\documentclass[preprint]{elsarticle}

\usepackage{lineno,hyperref}
\modulolinenumbers[1]

\usepackage{lipsum}
\usepackage{rotating}
\graphicspath{{figures/}}
\usepackage{pstricks, pst-node, psfrag}
\usepackage{amssymb,amsmath}
\usepackage{mathtools}
\usepackage{verbatim,enumerate}
\usepackage{rotating, lscape}
\usepackage{setspace}
\usepackage[font=small,labelfont=bf]{caption}
\usepackage[hang, flushmargin]{footmisc}
\usepackage{subfig}
\usepackage{caption}
\usepackage{cancel}
\usepackage{float}
\usepackage{ragged2e}
\usepackage{centernot}
\usepackage{tikz}
\usetikzlibrary{shapes,arrows,positioning}
\usepackage{textcomp} 
\usepackage{gensymb} 

\tikzstyle{endpt} = [rectangle, draw, fill=red!20,
    text width=9.8em, text centered, rounded corners, minimum height=4em]
\tikzstyle{block} = [rectangle, draw, top color=white, bottom color=blue!20,
    text width=9.8em, text centered, rounded corners, minimum height=4em]
\tikzstyle{line} = [draw, -latex', very thick]

\usepackage{listings}
\usepackage{color} 
\definecolor{mygreen}{RGB}{11,102,35}
\definecolor{mylilas}{RGB}{170,55,241}
\definecolor{mygray}{rgb}{0.5,0.5,0.5}

\definecolor{backgreen}{rgb}{0.00, 0.169, 0.212}
\definecolor{textgray}{rgb}{0.514, 0.580, 0.589}

\usepackage[T1]{fontenc}
\newcommand*\patchAmsMathEnvironmentForLineno[1]{%
  \expandafter\let\csname old#1\expandafter\endcsname\csname #1\endcsname
  \expandafter\let\csname oldend#1\expandafter\endcsname\csname end#1\endcsname
  \renewenvironment{#1}%
     {\linenomath\csname old#1\endcsname}%
     {\csname oldend#1\endcsname\endlinenomath}}%
\newcommand*\patchBothAmsMathEnvironmentsForLineno[1]{%
  \patchAmsMathEnvironmentForLineno{#1}%
  \patchAmsMathEnvironmentForLineno{#1*}}%
\AtBeginDocument{%
\patchBothAmsMathEnvironmentsForLineno{equation}%
\patchBothAmsMathEnvironmentsForLineno{align}%
\patchBothAmsMathEnvironmentsForLineno{flalign}%
\patchBothAmsMathEnvironmentsForLineno{alignat}%
\patchBothAmsMathEnvironmentsForLineno{gather}%
\patchBothAmsMathEnvironmentsForLineno{multline}%
}

\newcommand{\R}{\mathbb{R}}

\newcommand{\cG}{\mathcal{G}}

\newcommand{\cN}{\mathcal{N}}
\newcommand{\cO}{\mathcal{O}}
\newcommand{\cU}{\mathcal{U}}

\newcommand{\CC}{C\nolinebreak\hspace{-.05em}\raisebox{.4ex}{\tiny\bf +}\nolinebreak\hspace{-.10em}\raisebox{.4ex}{\tiny\bf +}}

\newcommand{\pder}[2][]{\frac{\partial#1}{\partial#2}}

\newcommand{\dt}{\Delta t}
\newcommand{\dx}{\Delta x}

\renewcommand{\vec}[1]{\boldsymbol{#1}}
\renewcommand{\bar}{\overline}
\renewcommand{\tilde}{\widetilde}
\renewcommand{\hat}{\widehat}

\newcommand{\defeq}{\coloneqq}

\DeclareMathOperator\diag{diag}

\DeclareMathOperator\RMSE{RMSE}

\newcommand{\revone}[1]{\textcolor{black}{#1}}
\newcommand{\revtwo}[1]{\textcolor{black}{#1}}

\journal{Journal of Computational Physics}

\begin{document}

\begin{frontmatter}

\title{A Lagrangian Method for Reactive Transport with Solid/Aqueous Chemical Phase Interaction \tnoteref{mytitlenote}}
\tnotetext[mytitlenote]{The authors were supported by the National Science Foundation under awards EAR-1417145 and DMS-1614586.}

\author{Michael J. Schmidt\fnref{hydro,ams}}
\ead{mschmidt1@mines.edu}
\author{Stephen D. Pankavich\fnref{ams}}
\ead{pankavic@mines.edu}
\author{Alexis Navarre-Sitchler\fnref{hydro}}
\ead{asitchle@mines.edu}
\author{David A. Benson\fnref{hydro}}
\ead{dbenson@mines.edu}
\address{Colorado School of Mines\\ 1500 Illinois St.\\ Golden, CO 80401}
\fntext[hydro]{Hydrologic Science and Engineering Program, Department of Geology and Geological Engineering, Colorado School of Mines, Golden, CO, 80401, USA}
\fntext[ams]{Department of Applied Mathematics and Statistics, Colorado School of Mines, Golden, CO, 80401, USA}

\begin{abstract}

A significant drawback of Lagrangian (particle-tracking) reactive transport models has been their inability to properly simulate interactions between solid and liquid chemical phases, such as dissolution and precipitation reactions.
This work addresses that problem by implementing a mass-transfer algorithm between mobile and immobile sets of particles that allows aqueous species of reactant that are undergoing transport to interact with stationary solid species.
This mass-transfer algorithm is demonstrated to solve the diffusion equation for an arbitrarily small level of diffusion and thus does not introduce any spurious mixing.
The algorithm can be combined with random walks to simulate the desired total level of diffusion in a reactive transport system.

\end{abstract}


\end{frontmatter}


\section{Introduction} 
\label{sec:introduction}

Reactive transport models fall into two main categories: Eulerian (grid-based) models (e.g., finite difference, finite volume, and finite element) or Lagrangian models (e.g., particle-tracking (PT)).
For a comparison between Eulerian and Lagrangian reactive transport methods, in terms of accuracy and computational gains, see \cite{Benson_AWR_2016}.
Eulerian models, as used in research and industry, are far more common, but, for certain problems, they are known to suffer from important drawbacks due to their use of a static grid.
Namely, concentration fluctuations occurring at a length scale smaller than that of the grid are not captured.
This imposes a cutoff point for the model, such that heterogeneities and fluctuations may be resolved between grid elements, but concentrations are perfectly-mixed within each grid element.
Similarly, if an Eulerian model user wishes to capture smaller-scale fluctuations, then a more finely discretized grid must be employed, leading to longer computation times.
This discretization-dependent homogeneity assumption has important implications for the modeling of chemical reactions undergoing transport via diffusion and/or advection.
The traditional way to handle this has been by empirical adjustments to reaction rate constants \cite{battiato1,battiato2,Schwede_sample_pdf}.
Another known shortcoming of Eulerian methods is their introduction of numerical diffusion in the simulation of advection \cite{Sweby1984,Leonard1991,Leveque2002}, a phenomenon which does not occur in PT methods.

In contrast, Lagrangian reactive transport models do not use a static grid that reactants move through.
Instead, concentrations (or masses) of reactant are carried with the elements (particles) as the particles, themselves, move through the computational domain.
For this reason, particle methods impose no cutoff resolution and can capture heterogeneities and fluctuations on a fully continuous scale.
This may be achieved both by allowing for mass differences between particles \cite[e.g.,][]{Bolster2016,Benson_arbitrary}, and letting the spatial distribution of particles change with time.
The heterogeneity captured by inter-particle spacings is related to the level of discretization (i.e., particle number), and this method of representing spatial fluctuations of the concentrations is not possible with Eulerian methods.
Another valuable property of these PT algorithms is that systems with a higher level of concentration heterogeneity (i.e., high spatial variance in concentration) may be modeled using fewer particles than would be required for a more homogeneous (or ``smoother'') system, as the number of particles is directly tied to a system's degree of spatial concentration variability \cite{Paster_JCP}.

However, this efficiency in capturing heterogeneity can also be a disadvantage when one needs to model a system including a very smooth, or homogeneous, concentration field.
In order to properly resolve a very smooth field, classic particle methods would require a very large number of particles, leading to high computational burden (number of particles is analogous to number of grid points in an Eulerian model).
Two methods for addressing this problem have recently been proposed.
First, a spatial kernel with nonzero support may be employed (a Gaussian kernel is the standard choice, due to its natural compatibility with the diffusion process), in contrast to the classic Dirac delta kernel.
This allows for significant reduction in particle number while introducing a controllably low level of error \cite{schmidt2017}.
Alternatively, for complex reactions involving many chemical species, particle number may be reduced by allowing each particle to carry an arbitrary number of different species (rather than a single species per particle) that are transferred between/among particles according to a diffusive mass-transfer algorithm \cite{Benson_arbitrary,Engdahl_WRR,mass_trans_acc}.

Another shortcoming of Lagrangian PT methods, which we address in this paper, is their inability to model interactions between solid and aqueous chemical phases (e.g., dissolution or precipitation reactions, wherein flowing liquid carrying aqueous chemicals interacts with the surrounding rock).
The difficulty of this problem lies at the core of the Lagrangian framework because the particles in a PT simulation move, but stationary solids must not.
A na\"ive approach to this problem may employ a ``reaction grid'' \revtwo{through which particles carrying aqueous species move}, and any particles within a given grid element are considered to be a well-mixed reaction volume.
However, this type of approach imposes the previously-described over-mixing problem.
Thus, we propose a framework in which mobile particles (carrying aqueous chemical phases) interact with immobile particles (carrying solid chemical phases) through a diffusive mass-transfer (MT) process that imposes no spurious mixing.

First, in Section \ref{sec:analytic_model_and_methods}, we define the reactive transport problem considered herein and describe the mobile-immobile reactive particle-tracking (miRPT) algorithm used to model the problem.
In Section \ref{sub:particle_tracking_model}, the mass-transfer algorithm that is central to the miRPT model is derived in detail, and computational considerations necessary for simulating such a model are presented.
We derive a ``diffusion operator'' in Section \ref{sub:derivation_of_diffusion_operator} that is a discretized Green's function solution and use the solutions given by this diffusion operator for comparison with the miRPT MT algorithm throughout Section \ref{sec:results}.
In Section \ref{sub:analysis_of_mobile_immobile_mass_transfer_algorithm}, we first examine the behavior of the miRPT MT algorithm in isolation (without reaction) and demonstrate that it solves the diffusion equation with a controllable level of accuracy.
We also examine the effect of the various discretization parameters (number of mobile or immobile particles and time step length) on the accuracy of the miRPT MT algorithm and describe a stability condition that should be taken into consideration when implementing the algorithm.
Finally, in Section \ref{sub:reactive_transport_model}, we discuss the results of using the miRPT algorithm to model a chemically-complex reactive transport model that involves calcite dissolution and dolomite precipitation/dissolution.
We describe the process of choosing the correct discretization parameters for this system in \ref{sub:selection_of_model_parameters}, and we give some important computational details of the model in \ref{sub:computational_details_dolo}.



\section{Analytic model and methods} 
\label{sec:analytic_model_and_methods}

\subsection{Governing equations} 
\label{sub:governing_equations}

We consider a reactive system of arbitrary chemical complexity that is undergoing transport by advection and diffusion.
For an infinite, $d$-dimensional domain ($d = 1, 2, 3$), the governing (system of) equation(s) for this reactive transport system is the advection-diffusion-reaction equation (ADRE)
\begin{gather}\label{adre}
    \pder[C_i]{t} + \nabla \cdot \left(v_i C_i - D_i \nabla C_i\right) = r_i(C_1, \dots, C_n, k_1, \dots, k_m),\\
    i = 1, \dots, n, \quad x \in \R^d, \quad t > 0, \nonumber
\end{gather}
where the concentration of species $i$ is given by $C_i(t,x)$ [mol L$^{-d}$], $D_i$, $v_i$, and $k_j$ are the diffusion coefficient [L$^2$ T$^{-1}$], velocity [L T$^{-1}$], and reaction rate constant [L$^d$ mol$^{-1}$ T$^{-1}$], respectively, for species $i$ and reaction channel $j = 1, \dots, m$.
The reaction term, $r_i(\cdot, \dots, \cdot)$ is some, possibly nonlinear, function assumed to be governed by the law of mass action.
We note that for a solid species (not undergoing transport) $v = D \equiv 0$, as concentrations will only be altered by (dissolution or precipitation) reactions.
As well, for the purposes of this paper, we assume that, for aqueous species, $v$ and $D$ are constant in both space and time.


\subsection{Particle-tracking model} 
\label{sub:particle_tracking_model}

\emph{Benson and Meerschaert} \cite{Benson_react} developed their reactive particle-tracking (RPT) method in order to simulate chemical reaction in diffusive media.
The RPT algorithm models reactions by calculating the probability of reaction between particle pairs that is conditioned upon the event of particles co-locating by molecular diffusion.
Under this paradigm, each particle carries a given mass of only one type of reactant.
When a pair of particles react they are ``killed,'' converting the mass that they carry to product, and the mass of the corresponding species are thus altered.
This approach has been extended beyond the particle-killing approach to assign continuously-varying masses to the particles that are reduced \emph{in proportion to} their co-location probability \cite{Bolster_mass}.
It has also been extended to allow individual particles to carry multiple species of reactant (obviating the need for a distinct, large set of particles for each individual species) and exchange mass among one another, thus allowing reactions to occur \emph{within} a given particle that is treated as a well-mixed reaction volume \cite{Benson_arbitrary}.

It is this final, arbitrarily-complex reactive particle-tracking (aRPT) algorithm that we base our work on in this paper.
It was recently shown that the mass-transfer (MT) process employed by aRPT solves the diffusion equation with $\cO(\dt)$ accuracy \cite{mass_trans_acc}.
Thus, the total diffusion in a system may be partitioned between the random walk and MT numerical processes with a controllable level of error.
\emph{Benson and Bolster} \cite{Benson_arbitrary} attempted to simulate mobile-immobile particle interaction with a na\"ive implementation of the aRPT algorithm that resulted in large errors, attributable to excess numerical diffusion.
However, we modify this MT algorithm to handle interactions between aqueous and solid (mobile/immobile) phases of reactant and thus formulate our approach to a mobile-immobile RPT (miRPT) algorithm in the following section.
Additionally, we show that the altered MT algorithm also solves the diffusion equation with minimal error.

\subsubsection{Mobile-immobile reactive particle-tracking} 
\label{ssub:mobile_immobile_reactive_particle_tracking}

\begin{figure}[tp]
    \centering
    \begin{tikzpicture}[node distance = 2.5cm, auto]
        \node [endpt] (init) {Initial masses (mobile/immobile)};
        \node [anchor=north west,font=\tiny\color{black}] at (init.north west) {(a)};
        \node [block, below of=init] (trans) {\ \ Transport particles\\ (random walk \& advection, Eq. \eqref{transport_langevin})};
        \node [anchor=north west,font=\tiny\color{black}] at (trans.north west) {(b)};
        \node [block, below of=trans] (mob2im) {Transfer mobile species masses to immobile particles ($\vec W_M \vec m_M = \vec m_I$, Eq. \eqref{matmul})};
        \node [anchor=north west,font=\tiny\color{black}] at (mob2im.north west) {(c)};
        \node [block, right of = mob2im, node distance = 4.2cm] (mass2conc) {Convert masses to concentrations};
        \node [anchor=north west,font=\tiny\color{black}] at (mass2conc.north west) {(d)};
        \node [block, above of=mass2conc] (rxn) {Perform reactions\\ on immobile particles};
        \node [anchor=north west,font=\tiny\color{black}] at (rxn.north west) {(e)};
        \node [block, above of=rxn] (conc2mass) {Convert concentrations to masses};
        \node [anchor=north west,font=\tiny\color{black}] at (conc2mass.north west) {(f)};
        \node [block, right of=conc2mass, node distance = 4.2cm] (im2mob) {Transfer mobile species masses to mobile particles ($\vec W_I \vec m_I = \vec m_M$, Eq. \eqref{matmul})};
        \node [anchor=north west,font=\tiny\color{black}] at (im2mob.north west) {(g)};
        \node [endpt, below of=im2mob] (final) {Final masses\\ (mobile/immobile)};
        \node [anchor=north west,font=\tiny\color{black}] at (final.north west) {(h)};
        \path [line] (init) -- (trans);
        \path [line] (trans) -- (mob2im);
        \path [line] (mob2im) -- (mass2conc);
        \path [line] (mass2conc) -- (rxn);
        \path [line] (rxn) -- (conc2mass);
        \path [line] (conc2mass) -- (im2mob);
        \path [line] (im2mob) -- (final);
        \path[line] (final) -| ([xshift=0.3cm,yshift=0.3cm]im2mob.north east) -| node [near start, above] {\textbf{Step in time}} (init);
    \end{tikzpicture}
    \caption{Flowchart demonstrating the order of operations for the miRPT algorithm.}
    \label{fig:imRPT_chart}
\end{figure}
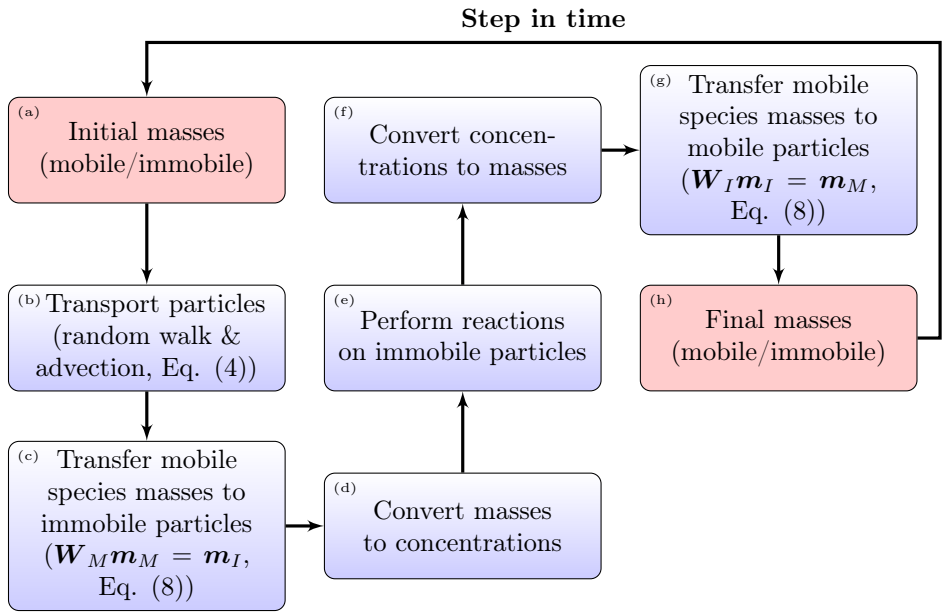

The essential motivation behind the miRPT model is that the algorithm employs two sets of particles (mobile and immobile) because aqueous and solid species (henceforth referred to as mobile/immobile mass, respectively, within the context of the algorithm) must interact with one another, but solid species must not be transported.
As a result, all chemical reaction calculations are performed by treating the immobile particles as a well-mixed volume and allowing the mobile mass to be transferred between the mobile and immobile particles within a time step according to an MT algorithm that is shown to be diffusive.
The size of this well-mixed volume may be defined by the user with the chemical reactions in mind, as opposed to, in the Eulerian paradigm, choosing the well-mixed volume size based on constraints imposed by the transport algorithm (i.e., bounds on the P\'eclet or Courant number).
A step-by-step outline of the miRPT algorithm is given by the flowchart in Figure \ref{fig:imRPT_chart}.

We specify here that, for the miRPT algorithm described in this paper, \emph{all} of the mobile mass is transfered to the immobile particles for reaction and back to the mobile particles for transport (steps (c) and (g) in Figure \ref{fig:imRPT_chart}).
\revtwo{The reason for this is that chemical reactions are performed on the immobile particles, and all mass that is eligible for reaction must be moved there via mass transfer.
This implies that there must be a sufficiently dense spatial distribution of immobile particles in the domain so as not to introduce error in the simulation of diffusion.
This constraint leads to the stability condition discussed in Section \ref{sub:analysis_of_mobile_immobile_mass_transfer_algorithm}.
}
Presumably, this constraint could be relaxed by performing reaction calculations for interactions \emph{between} mobile and immobile particles; however, this would require precise control over both the reaction calculations and over which reactions are calculated between certain particles.
For this paper, we wish to consider the most general case, wherein all reactions are calculated simultaneously among all species of reactant.
This treatment allows for the use of geochemical solvers such as PHREEQC (PhreeqcRM), CrunchFlow, or Reaktoro \cite{phreeqcrm,CRUNCHFLOW,leal2015reaktoro}.

\subsubsection{Algorithm details} 
\label{ssub:algorithm_details}

Particle-tracking algorithms represent concentrations by employing a spatial kernel. The most common choice of kernel is a Dirac delta, though other choices have been shown to be effective, particularly if a smaller number of particles is desired for a given initial condition \cite{schmidt2017, guillem2017kde}.
We note here the distinction that, due to the infinitesimal support of a Dirac delta, particles typically carry \emph{masses}, as opposed to the \emph{concentrations} tracked at grid locations of analogous Eulerian methods.
However, concentrations may be easily generated by grouping particles into ``bins'' of fixed volume (in this case, the immobile particles represent a fixed volume).

Concentrations, in phase $p = M, I$ (mobile or immobile), are represented in space in the following way
\begin{equation}\label{genconc}
    \begin{aligned}
        C_p(t, x) &= \int_\Omega \sum_{j = 1}^{N_p} m_j^{(p)} \phi(x - z)\delta\left(z - x_j^{(p)}\right) dz, \quad p = M, I, \\
        &= \sum_{j = 1}^{N_p} m_j^{(p)} \phi\left(x - x_j^{(p)}\right),
    \end{aligned}
\end{equation}
where $\Omega$ is the $d$-dimensional domain ($d = 1, 2, 3$), $m_j^{(p)}(t)$ and $x_j^{(p)}(t)$ are the mass and position of the $j^{\text{th}}$ particle in phase $p$ (the time-dependence of each of these is suppressed above), $N_p$ is the number of particles of phase $p$, $\phi$ is the spatial kernel (assumed to be symmetric and integrate to unity), and $\delta$ is a Dirac delta.
For the specific choice of Dirac delta spatial kernel (as in this work), \eqref{genconc} becomes
\begin{equation}\label{delta_conc}
    C_p(t, x) = \sum_{j = 1}^{N_p} m_j^{(p)} \delta\left(x - x_j^{(p)}\right).
\end{equation}

The positions of mobile particles evolve according to the stochastic Langevin equation, as derived from Fick's Law
\begin{equation}\label{transport_langevin}
    x_j^{(M)}(t + \dt) = x_j^{(M)}(t) + \vec \xi_j \sqrt{2 D \dt} + v \dt,
\end{equation}
where $\vec \xi_j$ is a $d$-dimensional $(d = 1, 2, 3)$ random vector, and each component is an independent standard normally-distributed $\left(\cN(0, 1)\right)$ random variable.

We formulate the miRPT MT algorithm as a generalization of the work of \emph{Benson and Bolster} \cite{Benson_arbitrary} such that transfer of mobile mass from one phase to another may be formulated as a weighted sum of mass exchanges
\begin{equation}\label{ctx}
    \begin{aligned}
        C_{\hat p} (t, x) &= \sum_{j = 1}^{N_{p}} \left[ m_j^{(p)} \sum_{k = 1}^{N_{\hat p}} w^{(p)}\left(x_k^{(\hat p)} ; x_j^{(p)}\right) \delta\left(x - x_k^{(\hat p)}\right) \right] \\
        &= \sum_{k = 1}^{N_{\hat p}} m_k^{(\hat p)} \delta\left(x - x_k^{(\hat p)}\right), \quad p, \hat p = M, I,\quad p \neq \hat p,
    \end{aligned}
\end{equation}
where $w^{(p)}\left(x_k^{(\hat p)} ; x_j^{(p)}\right)$ is the weighting function for the mass transfer from the $j^{\text{th}}$ particle in phase $p$ to the $k^{\text{th}}$ particle in phase $\hat p$ and need not be the same for $p = I$ or $M$.
Since all mobile mass is transferred from one phase to the other by the miRPT algorithm, this implies that the weights sum to 1 over the $p$ index, i.e.,
\begin{equation}\label{wt_sum}
    \sum_{j = 1}^{N_{p}} w^{(p)}\left(x_k^{(\hat p)} ; x_j^{(p)}\right) = 1,
\end{equation}
or,
\begin{equation}\label{matmul_sum}
    \sum_{j = 1}^{N_{p}} m_j^{(p)} w^{(p)}\left(x_k^{(\hat p)} ; x_j^{(p)}\right) = m_k^{(\hat p)}.
\end{equation}
We see that the form of the sum in \eqref{matmul_sum} implies that this may be written in matrix-vector form as
\begin{equation}\label{matmul}
    \vec W_p \vec m_p = \vec m_{\hat p},
\end{equation}
where the entries of the matrix and vectors are $\left[\vec W_p\right]_{kj} = w^{(p)}\left(x_k^{(\hat p)} ; x_j^{(p)}\right)$ and $\left(\vec m_z\right)_i = m_i^{(z)},\ z = p, \hat p$.

The specific weighting function we use for miRPT is based on the probability density for the co-location of a mobile-immobile particle pair by diffusion within a timestep of length $\dt$ \cite{Benson_react}
\begin{equation}\label{co_loc_density}
    v_{\kappa^{(p)}}(s)=\frac{1}{\left(\kappa^{(p)} 4\pi D\Delta t\right)^{d/2}}\exp\left[{-\frac{\vert s\vert^2}{\kappa^{(p)} 4 D\Delta t}}\right],
\end{equation}
where $d = 1, 2, 3$, is the number of spatial dimensions, $D$ is the diffusion coefficient, $\vert s \vert$ is the (Euclidean) distance between the pair of particles, and $\kappa^{(p)}$ is a scaling constant on the portion of diffusion to be simulated by a single mass transfer that may be different for $p = M, I$\revone{, provided that $\kappa^{(M)} + \kappa^{(I)} = 1$ and $\kappa^{(p)} \neq \{0, 1\}$ (the end member cases are excluded, as under this framework no mass transfer would occur).
A mathematical justification for this partitioning of diffusion is discussed in \ref{sub:justification_for_partitioning_diffusion_via_}.
However, from a physical standpoint, $\kappa^{(p)}$ encodes the magnitude of diffusive mass transfer from the mobile phase to the immobile phase, or vice versa (this can equivalently be viewed as the magnitude of spatial variance added to the solute plume by a given transfer).
For example, holding $\kappa^{(M)} = \kappa^{(I)} = 0.5$ implies that mobile-to-immobile and immobile-to-mobile transfers will simulate an equal amount of the total diffusion, whereas any other combination implies unequal amounts of diffusion in each ``direction'' of mass-transfer.
}

Now, in order to enforce \eqref{wt_sum}, we select $w^{(p)}$ to be
\begin{equation}
    w^{(p)} \left(x_k^{(\hat p)} ; x_j^{(p)}\right) = v_{0.5}\left(x_j^{(\hat p)} - x_k^{(p)}\right)\left[\sum_{\ell = 1}^{N_p} v_{0.5}\left(x_j^{(\hat p)} - x_{\ell}^{(p)}\right)\right]^{-1}.
\end{equation}
The reason for this choice is that we would like the miRPT MT process to be diffusive, as was shown for a similar MT algorithm in \cite{mass_trans_acc}, and division by the sum term in brackets ensures that the weights sum to unity and all mass is transferred from one phase to the other (computational details for this process are discussed in \ref{ssub:computational_details_for_the_mass_transfer_process}).
We also choose $\kappa^{(M)} = \kappa^{(I)} = 0.5$, and this choice is examined in \ref{sub:relationship_between_error_and_kappap_}.

As a side note, the reader may notice that $v_2(s)$ would be the co-location probability density for a mobile-mobile collision.
Additionally, if, in \eqref{matmul}, we treat phase $p$ as mobile particles at $t = t_0$ and $\hat p$ as mobile particles at $t = t_0 + \dt$ (this requires that $N_p \equiv N_{\hat p}$), we may use this density to define the weight matrix
\begin{equation}
    \vec W_E \defeq \vec I + \frac12 \Big[\vec P - \diag(\vec P \vec 1)\Big],
\end{equation}
where $\left[\vec P_{kj}\right] = v_2\left(x_j^{(\hat p)} - x_k^{(p)}\right)$, $\diag(\vec x)$ denotes a diagonal matrix with the entries of the vector $\vec x$ on the main diagonal, and $\vec 1$ is a $N \times 1$ vector of ones.
This is the matrix associated with the ``explicit matrix'' method discussed in \cite{mass_trans_acc}.
That is to say, the miRPT method recovers the classical mobile-mobile mass-transfer algorithm.



\subsubsection{Computational details} 
\label{ssub:computational_details}

In this section, we discuss details that are not essential to the reactive transport portion of the algorithm but are nonetheless necessary from a computational/coding standpoint.
The first concern to address is the (typically) computationally expensive nature of chemistry calculations.
The miRPT algorithm has a natural ability to handle this, as all of the chemistry calculations are performed ``on'' the immobile particles, and the number of immobile particles $\left(N_I\right)$ is a parameter that may be adjusted according to the desired balance between error and computational expense.
As we show in Section \ref{par:relationship_between_error_and_Nm/Ni}, the error in the miRPT algorithm as a diffusive operator is at its lowest when the number of immobile particles is greater than or equal to the number of mobile particles $\left(N_I \geq N_M\right)$; however, $N_I$ may be adjusted downward $\left(\text{e.g., }N_I = N_M / 2\right)$ in favor of conducting fewer chemistry calculations, and the aforementioned error may still be made relatively small for appropriate choices of other simulation parameters $\left(\dt,\ N_M\right)$.

From a theoretical standpoint, according to the miRPT MT algorithm outlined in Section \ref{ssub:algorithm_details}, every particle exchanges mass with every other particle, regardless of separation distance.
However, because the co-location probability density given in \eqref{co_loc_density} is Gaussian, we know that particles with a separation distance larger than three standard deviations $\left( \sigma = \sqrt{\kappa^{(p)} 4 D \dt}\right)$ have $\lessapprox 0.3\%$ probability of co-locating and exchanging mass.
For this reason, a cutoff distance for MT interactions is typically imposed, so that particles with a separation distance greater than $3\sigma$ do not interact.
If this cutoff distance approach is combined with a linear algebra package that takes advantage of sparse matrix structure, this can provide substantial computational speedup and memory savings.
As well, imposing this cutoff distance lends the problem to highly efficient fixed-radius search algorithms $\big(\cO\left(N \log N\right) \text{, as compared to } \cO\left(N^2\right)$ for a na\"ive search$\big)$, where the standard choice is a kD Tree search algorithm \cite{og_kd} such as MATLAB's \texttt{rangesearch()} \cite{matlab_kd} or an open-source version available in \CC\ or Fortran 95 \cite{kennel}.
Finally, all of the above computational strategies become essential for models with higher than one spatial dimension, as dense matrices of the required size often will not fit in memory, and a na\"ive distance search quickly becomes computationally intractable.



\subsection{Derivation of diffusion operator} 
\label{sub:derivation_of_diffusion_operator}

In order to analyze the accuracy of the miRPT MT algorithm of Section \ref{sub:analysis_of_mobile_immobile_mass_transfer_algorithm}, we wish to compare it to another numerical method that solves the continuum diffusion equation
\begin{equation}\label{diff_eqn}
    \pder[C]{t} = D \Delta C, \quad x \in \R^d, \quad t > 0,
\end{equation}
given a particle-like spatial discretization (i.e., diffusive mass transfers among only mobile particles).
\revtwo{Above, all quantities are as described in Section \ref{sub:governing_equations}, and $\Delta \equiv \nabla^2$ is the Laplacian.}
We outline the derivation of such a method here.

In an infinite $d$-dimensional domain, the solution to \eqref{diff_eqn} at time $T$ can be exactly expressed as a convolution of the initial condition (IC) with the associated Green's function.
Given $C_0(x) \defeq C(t = 0, x)$, and the Green's function for the diffusion equation $\cG(T, x) = \left(4 \pi D T\right)^{-d/2}\exp\left[-x^2/(4 D T)\right]$, we have
\begin{equation}\label{diff_conv}
    \begin{aligned}
        C(T, x) &= (\cG \star C_0) (T, x)\\
        &= \int_{\R^d} \cG(T, x - x_0) C_0(x_0) dx_0,
    \end{aligned}
\end{equation}
where $\star$ denotes convolution.
If we discretize time, such that $T = k \dt$, we may equivalently find our solution in the following way (employing pseudo-code)
\begin{equation}
    \begin{aligned}
        &\textbf{\texttt{for}}\ i = 1 : k\\
            &\qquad C(i \dt, x) = \int_{\R^d} \cG(\dt, x - x_0) C\big((i - 1) \dt, x_0\big) dx_0\\
        &\textbf{\texttt{end}}.
    \end{aligned}
\end{equation}
If we wish to use this same method to compute an exact solution in the case where our initial condition is composed of $N$ Dirac deltas, each with position $x_i$ and mass $m_i$ (an IC that corresponds to a particle-tracking simulation), we may insert the Dirac particle representation for $C(t = 0, x)$ from \eqref{delta_conc} into \eqref{diff_conv}
\begin{equation}\label{part_IC_cont}
    \begin{aligned}
        C(T, x) &= \int_{\R^d} \cG(T, x - x_0) \sum_{i = 1}^{N} m_i(t = 0) \delta(x_0 - x_i) dx_0\\
        &= \sum_{i = 1}^{N} m_i(t = 0) \cG(T, x - x_i).
    \end{aligned}
\end{equation}
Note, however, that this is a \emph{continuous} exact solution, given a particle-like IC.
If we wish our final solution to be represented by $N$ discrete particles, it must have the form
\begin{equation*}
    C(T, x) = \sum_{i = 1}^{N} m_i(t = T)\delta(x - x_i),
\end{equation*}
where
\begin{equation*}
    m_i(t = T) = \sum_{j = 1}^{N} m_j(t = 0) \cG\left(T, x_i - x_j\right),
\end{equation*}
or
\begin{equation}\label{Dtilde}
    \vec m(t = T) = \vec{\tilde D}_{T} \vec m(t = 0),
\end{equation}
where $\left[\vec{\tilde D}_{T} \right]_{ij} \defeq \cG\left(T, x_i - x_j\right)$ and $\left[\vec m(t)\right]_i \defeq m_i(t)$.
However, we see that this space-discretized diffusion operator, $\vec{\tilde D}_{T}$, does not conserve mass because each column is simply $N$ function evaluations from a Gaussian distribution function.
Thus, we must normalize the columns of $\vec{\tilde D}_{T}$ to sum to 1, which may be equivalently viewed as constructing a discrete probability mass function from the continuous distribution defined by $\cG$.
So, we define
\begin{equation}\label{D_from_tilde}
    \vec D_T \defeq \vec{\tilde D}_{T} \diag^{-1}\left(\vec 1 \vec{\tilde D}_{T}\right),
\end{equation}
where, as previously used, $\vec 1$ is a $1 \times N$ vector of ones, and $\diag^{-1}\left(\vec x\right)$ is a square matrix with the element-wise reciprocal of the vector $\vec x$ on its main diagonal.
So, substituting $\vec D_T$, above, into \eqref{Dtilde}, we have a representation for MT by a particle diffusion operator
\begin{equation}\label{diff_op_xfer}
    \begin{aligned}
        \vec m(t = T) &= \vec D_T \vec m(t = 0)\\
        &\approx \big[\vec D_{\dt} \big]^k \vec m(t = 0),
    \end{aligned}
\end{equation}
where the error in the approximation above is due to discretization and becomes equivalent to \eqref{part_IC_cont} as $N \to \infty$.

As discussed in \cite{mass_trans_acc}, the simulated solutions generated by this discretized matrix diffusion operator (henceforth referred to as the \emph{diffusion operator}) are nearly error-free when there is no influence from boundary conditions.
However, because the derivation assumes an infinite domain, error is introduced in a finite computational domain when particles near the boundaries have non-negligible mass.
This is due to the fact that the normalization in \eqref{D_from_tilde} alters the ``shape'' of the Green's function near the boundary.
Nonetheless, the diffusion operator serves as a good algorithmic benchmark for the miRPT MT algorithm if boundary effects are minimized.

\revone{We note here that the diffusion operator derived in this section is presented as a Green's function solution, discretized within a timestep on a mobile-particle ``grid.''
The miRPT MT algorithm uses the exact same approach for each direction of transfer.
However, the transfers are from one particle grid to another (i.e., the mobile grid to the immobile grid, or the reverse), and a link between the diffusion operator and the miRPT MT algorithm is presented in \ref{sub:justification_for_partitioning_diffusion_via_}.
Additionally, recent work has demonstrated that the type of mass-transfer algorithms presented in this text, as well as that in \cite{Benson_arbitrary} are part of a one-parameter family of smoothed-particle hydrodynamics (SPH) methods, allowing them to inherit a large body of theoretical underpinning \cite{guillem_SPH_equiv}.
}



\section{Results} 
\label{sec:results}

In this section, we show the results of testing the accuracy of the miRPT MT algorithm in solving the diffusion equation (Section \ref{sub:analysis_of_mobile_immobile_mass_transfer_algorithm}).
As well, we consider the results of using the full miRPT reactive transport algorithm to model a calcite/dolomite geochemical system in which CO$_2$-saturated brine is injected into a porous domain with solids composed of calcite and quartz (Section \ref{sub:reactive_transport_model}).
This problem is of interest because it involves both the dissolution of calcite and precipitation of dolomite \cite{Leal_2016_xLMA}, making it an ideal test problem for the miRPT algorithm.

Most numerical simulations in this section were conducted in MATLAB, using a MacBook Pro with a 2.9 GHz Intel Core i5 processor and 8 GB of RAM.
The more computationally-intensive MT-analysis simulations (2D and 100-member ensemble) were also conducted in MATLAB but were conducted on a 16-core node with two 2.7 GHz Intel Xeon E5 processors, with 8 cores each, and 64 GB of total RAM.
The full reactive transport simulation of the calcite/dolomite system was also run on the 16-core node, but code was written in Fortran 90 and employed the PhreeqcRM geochemical solver \cite{phreeqcrm} and kD Tree code from \cite{kennel}.
Additionally, we make all code used in this section available at \texttt{doi.org/10.5281/zenodo.2558584} \cite{miRPT_code}.

\subsection{Analysis of mobile-immobile mass-transfer algorithm} 
\label{sub:analysis_of_mobile_immobile_mass_transfer_algorithm}

We wish to gain an understanding of how the miRPT MT algorithm behaves in isolation and whether it solves the diffusion equation with a sufficient level of accuracy.
Therefore, we follow the approach of \emph{Schmidt et al.} \cite{mass_trans_acc} and consider only the MT algorithm, without other transport (random walks and advection) or reaction (i.e., simulating only steps (c) and (e) in Figure \ref{fig:imRPT_chart}).
Thus, we simulate a transfer of mass from the mobile phase to the immobile phase and then back to the mobile phase. Numerically, using \eqref{matmul}, this is represented by
\begin{equation}\label{wiwm}
    \vec W_I \vec W_M \vec m_M(t) = \vec m_M(t + \dt),
\end{equation}
where $\left[\vec m_M(t)\right]_i \defeq m_i^{(M)}(t)$. To asses the performance of the operator over the time period $T = k \dt$, we may apply the operator, $\big[\vec W_I \vec W_M\big]$, $k$ times to the initial mass vector as follows
\begin{equation}\label{miRPT_MT}
    \big[\vec W_I \vec W_M\big]^k \vec m_M(t = 0) = \vec m_M(t = T).
\end{equation}

We first consider the miRPT MT algorithm with equally-spaced particle positions and employ initial conditions (ICs) that have known analytic solutions, so that we may compare the results to these analytic solutions.
In addition, we compare these results to simulated solutions generated by the diffusion operator discussed in Section \ref{sub:derivation_of_diffusion_operator} that has been shown to yield near-exact solutions in the absence of influence from spatial boundaries \cite{mass_trans_acc}.
The two ICs chosen for this purpose are a domain-centered Gaussian (to avoid finite-domain boundary effects) and a Heaviside function (to investigate both the infinite gradient at a discontinuous interface and performance when boundary effects come into play).
We examine the sensitivity of the error between the analytic reference solution and the simulated solution generated by both miRPT MT and the diffusion operator.
This is done by successively refining the resolution provided by the parameter under consideration $\left(\dt,\ N_M,\ N_I\right)$ and tracking the resulting error, with the desired effect being that finer discretization will provide more accurate results.

Further, having found evidence of a stability condition for the miRPT MT numerical scheme, we investigate the influence of that condition.
\revtwo{The stability condition, defined and discussed subsequently (see \eqref{eta1}), is an empirically-motivated constraint on the miRPT MT algorithm that determines when it is capable of achieving an optimal level of error.
In many of the following error plots, the stability condition is plotted on the right-hand axis, and there is typically a distinct jump in error when a simulation enters or exits the region of stability (depicted by a solid black line).
}\
Also, we investigate whether the stability condition holds for a 2D simulation.

Lastly, we consider the miRPT MT algorithm's performance for randomly assigned particle positions.
In the case of the Gaussian IC, we first conduct a 100-member ensemble run and average the results.
We also conduct a single run and quantify the error by tracking the spatial variance in mass (given in Equation \eqref{spatial_variance}) in order to verify that it increases at the proper rate.
We also examine the variance increase for a non-analytic ``noisy box'' IC in the same manner.

We specify here that we include the diffusion operator approach as a comparison tool for the miRPT MT algorithm.
As is seen in the following sections, the diffusion operator approach achieves consistently lower error for many of the Gaussian IC cases, which is unsurprising since the diffusion operator is essentially interpolating a Gaussian solution using $N$ point-evaluations for each of $N$ Gaussian basis functions, a task for which it is naturally suited.
However, the diffusion operator is solving a different problem than the miRPT MT algorithm.
The diffusion operator approach is simulating diffusion with a discretized Green's function solution, using only the \emph{mobile} particles, i.e., (using \eqref{diff_op_xfer})
\begin{equation}
    \vec m_M(t = T) = \big[\vec D_{\dt} \big]^k \vec m_M(t = 0),
\end{equation}
whereas, the miRPT MT algorithm is simulating diffusion by performing the alternating mobile-immobile and immobile-mobile transfers (given in \eqref{miRPT_MT}) that are required for the more complicated reactive transport problem we are ultimately interested in simulating.
Thus, converging to the low error of the diffusion operator approach for a Gaussian IC is exactly the desired outcome, and outperforming the diffusion operator approach for the Heaviside IC, the 2D case, and all simulations with randomly-spaced particles is highly encouraging.

Unless specified otherwise, all simulations are conducted on a 1-dimensional domain, $\Omega = [0,1]$, for 1 second of simulation time and a diffusion coefficient of $D = 1.0 \times 10^{-3}$ m$^2$/s.
The parameter $\dt$ is chosen in order to properly explore the stability condition that is discussed below but typically varies between $1.0 \times 10^{-3}$ and $1.0 \times 10^{-1}$.
Further, the cutoff-distance approach described in Section \ref{ssub:computational_details} is not employed here (other than in the 2D case when memory becomes an issue), as computation time was not a concern for these simulations.

The measure of error used in these analyses will be root-mean-square error (RMSE), defined as follows for the RMSE between a given simulated solution $\vec C$ and reference solution $\vec C^*$ (both vectors of length $N$)
\begin{equation*}
    \RMSE\left( \vec C\right) \defeq \left[\frac{1}{N}\sum_{j = 1}^{N} \left(C^*_j - C_j\right)^2 \right]^{1/2}.
\end{equation*}

\subsubsection{Mass-transfer analysis for equally-spaced particles} 
\label{ssub:mass_transfer_analysis_for_equally_spaced_particles}

We first look to demonstrate that the miRPT MT algorithm solves the diffusion equation in the most idealized case, wherein particles are assigned fixed, equally-spaced positions in order to remove the randomness introduced by the diffusive random walks.
We stress here that using such a particle simulation to model the diffusion equation in practice would be non-optimal, as we have essentially recreated an Eulerian grid but are using an $N$-point stencil.
In this case, a finite difference (FD) or finite element method with a much narrower stencil would likely be a more computationally-efficient choice.
However, to reiterate, we do wish to show that the miRPT MT algorithm is diffusive so that we may partition a system's total diffusion between random walks and this MT process with a controllable level of numerical error.

In this section, convergence of the miRPT MT-simulated solutions to the analytic solution is investigated as it relates to all discretization parameters ($\dt,\ N_M,\ N_I$).
To do this, the parameter of interest is refined by successive halves to increase the resolution of the simulation, and we seek to verify that error is decreasing with the increased resolution.

\paragraph{Relationship between error and $\dt$} 
\label{par:relation_between_error_and_dt}

\begin{figure}[tp]%
    \centering
    \subfloat[Initial condition and final analytic and simulated solutions (plot shows simulated solutions for final data point, $\dt = 1/256$).]{\includegraphics[width=0.7\textwidth]{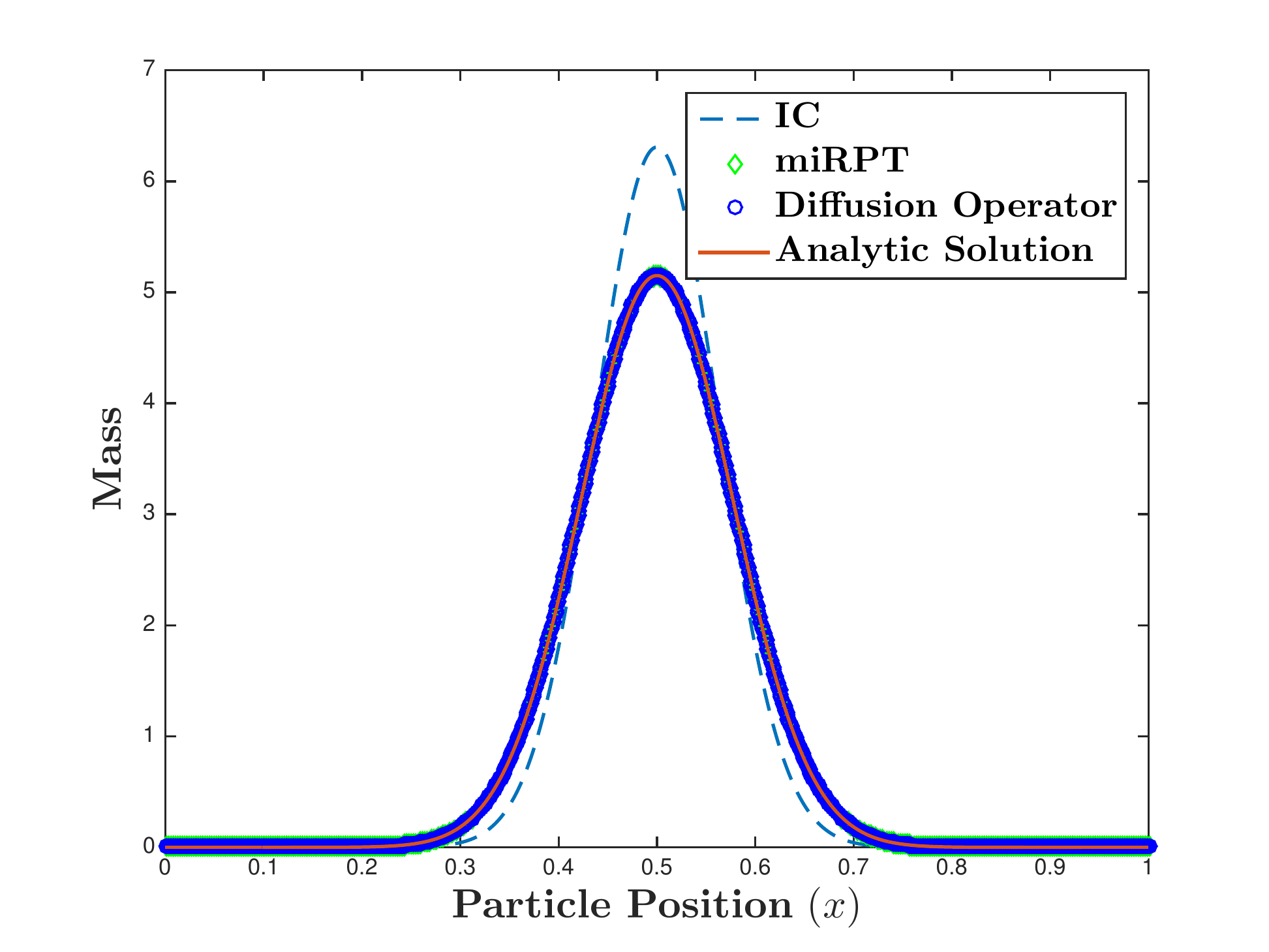} }
    \\
    \subfloat[RMSE vs. $1 / \dt$ (left axis) and stability condition (right axis).]{\includegraphics[width=0.7\textwidth]{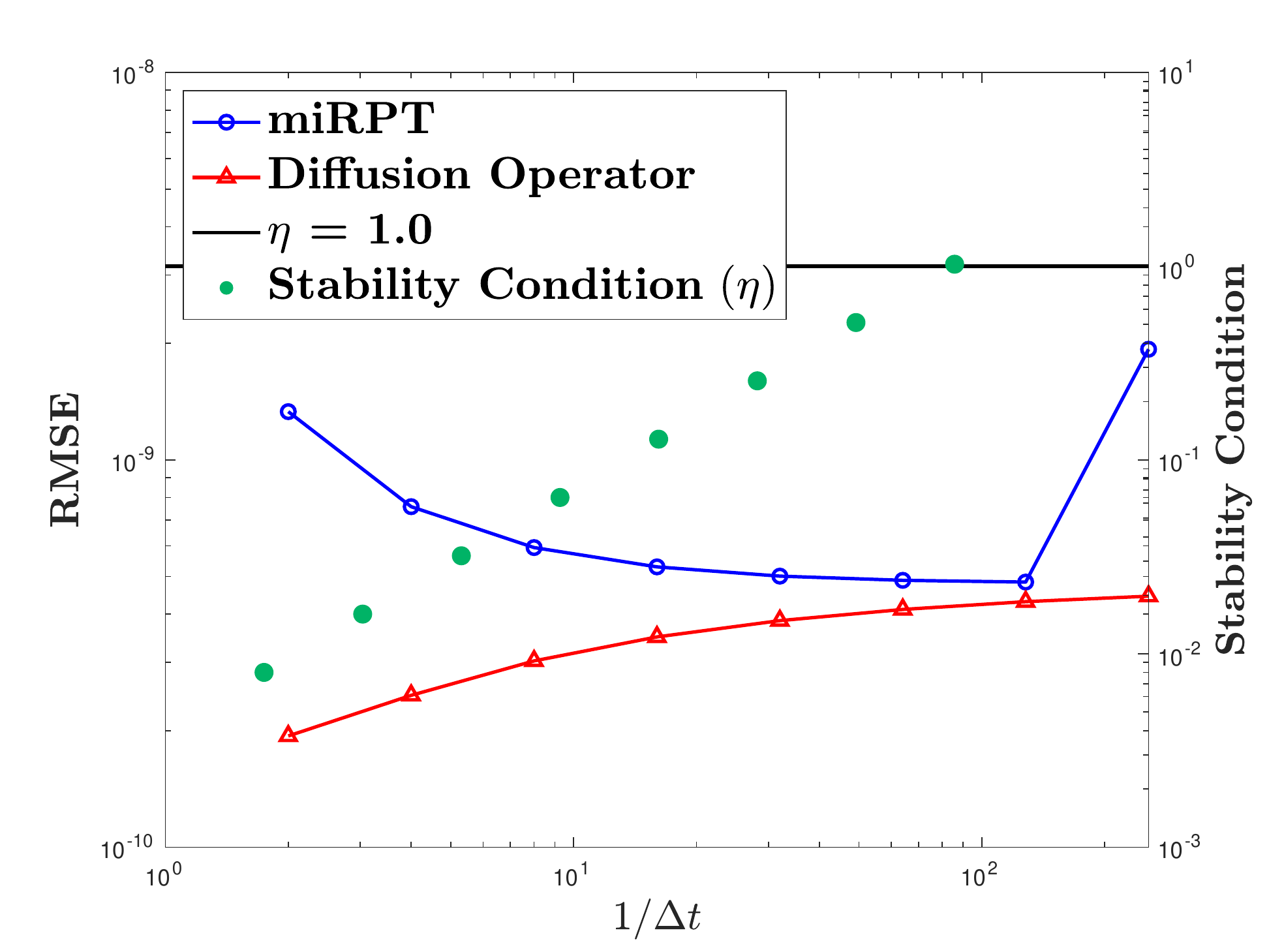} }
    \caption{Error/stability analysis as a function of $1 / \dt$ for Gaussian initial condition, with $N_I = 500,\ N_M = 1000$.}
    \label{fig:equi_dt_wGauss}
\end{figure}

\begin{figure}[tp]%
    \centering
    \subfloat[Initial condition and final analytic and simulated solutions (plot shows simulated solutions for final data point, $\dt = 1/256$).]{\includegraphics[width=0.7\textwidth]{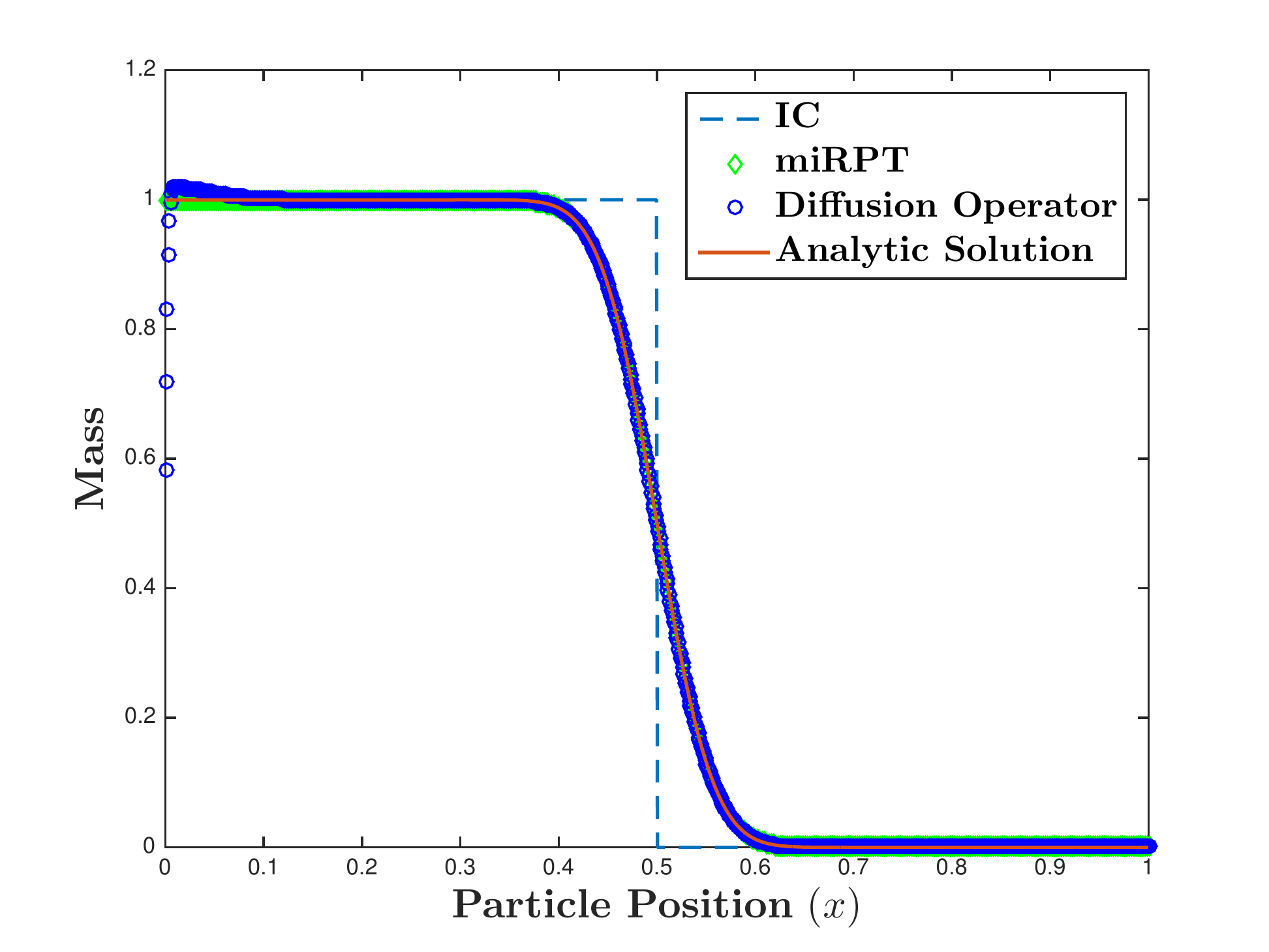} }
    \\
    \subfloat[RMSE vs. $1 / \dt$.]{\includegraphics[width=0.7\textwidth]{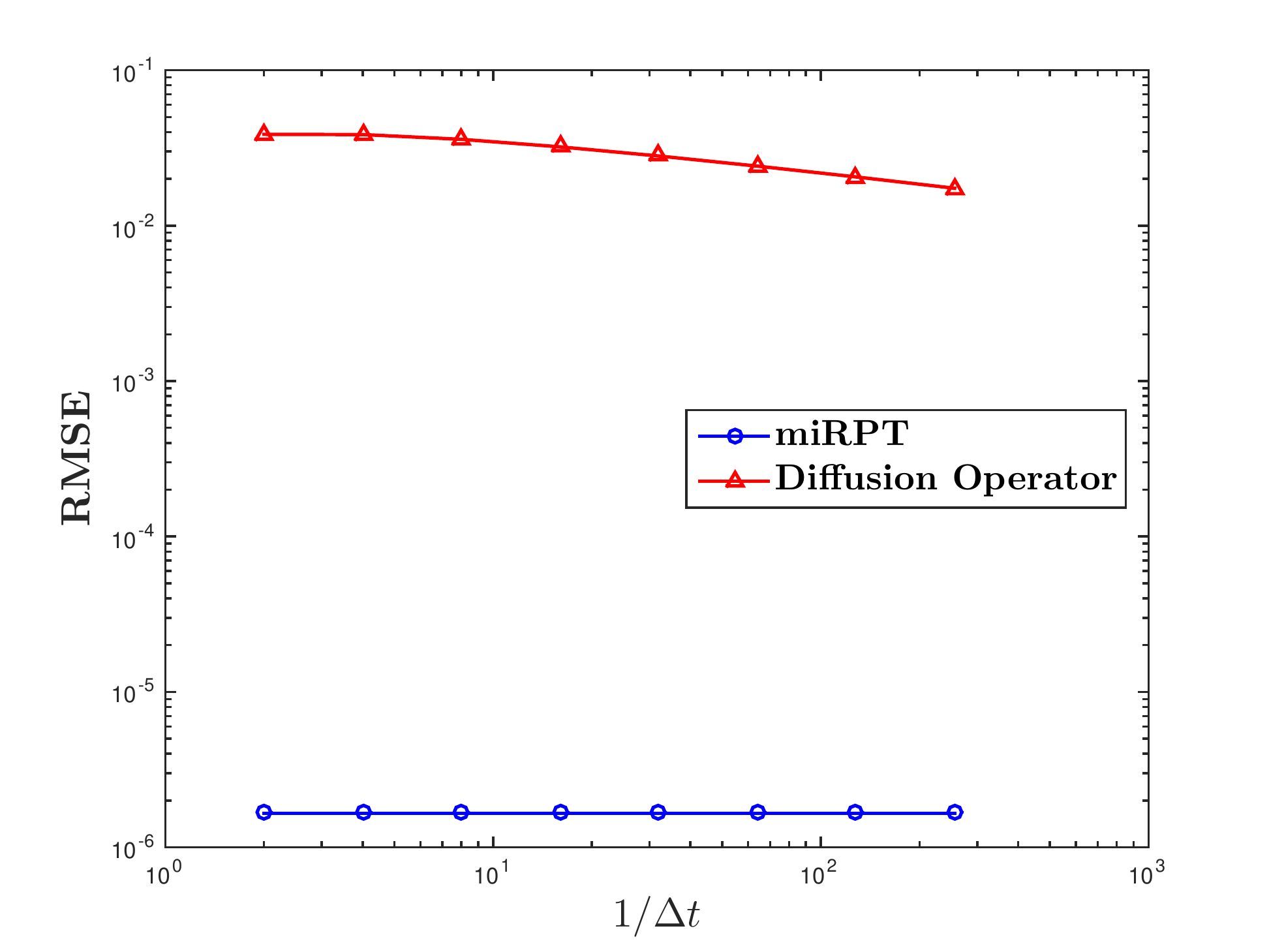} }
    \caption{Error analysis as a function of $1 / \dt$ for Heaviside initial condition, with $N_I = 500,\ N_M = 1000$.}
    \label{fig:equi_dt_heavi}
\end{figure}

Figures \ref{fig:equi_dt_wGauss} and \ref{fig:equi_dt_heavi} show the results of examining the sensitivity of error to the time discretization parameter, $\dt$.
Both figures demonstrate that the miRPT MT algorithm is relatively insensitive to changes in $\dt$ (as long as the stability condition discussed in a subsequent section is not violated).
For these simulations, $N_I = 500,\ N_M = 1000$.
We choose this ratio because, in a reactive transport model, it provides a desirable trade-off between minimizing the number of chemistry calculations to be performed (on the immobile particles) and minimizing the error due to the MT algorithm.
The effects of altering this ratio are explored in a following section.

In Figure \ref{fig:equi_dt_wGauss} (Gaussian IC), we consider the results when finite-boundary effects are minimized by employing the Gaussian IC and we see that error does, in fact, decrease with $\dt$.
As well, we see that the error in the miRPT MT algorithm appears to approach a minimal level of error that is achieved by the diffusion operator (whose error grows marginally as $\dt$ decreases).
In Figure \ref{fig:equi_dt_heavi} (Heaviside IC), we see a different story.
As discussed in \cite{mass_trans_acc}, the diffusion operator performs poorly when particles near the boundary have non-negligible mass, and that is clearly displayed in Figure \ref{fig:equi_dt_heavi}(a) where we see a vaguely Gaussian-shaped ``bump'' near the left-hand boundary, when those particles should all have a mass of exactly 1.
However, the miRPT MT algorithm attains an error many orders of magnitude smaller in this case, and hardly changes with refinements in $\dt$.


\paragraph{Relationship between error and particle number} 
\label{par:relationship_between_error_and_Nm}

\begin{figure}[tp]%
    \centering
    \subfloat[Initial condition and final analytic and simulated solutions (plot shows simulated solutions for final data point, $N_M = 3200 = 2 N_I$).]{\includegraphics[width=0.7\textwidth]{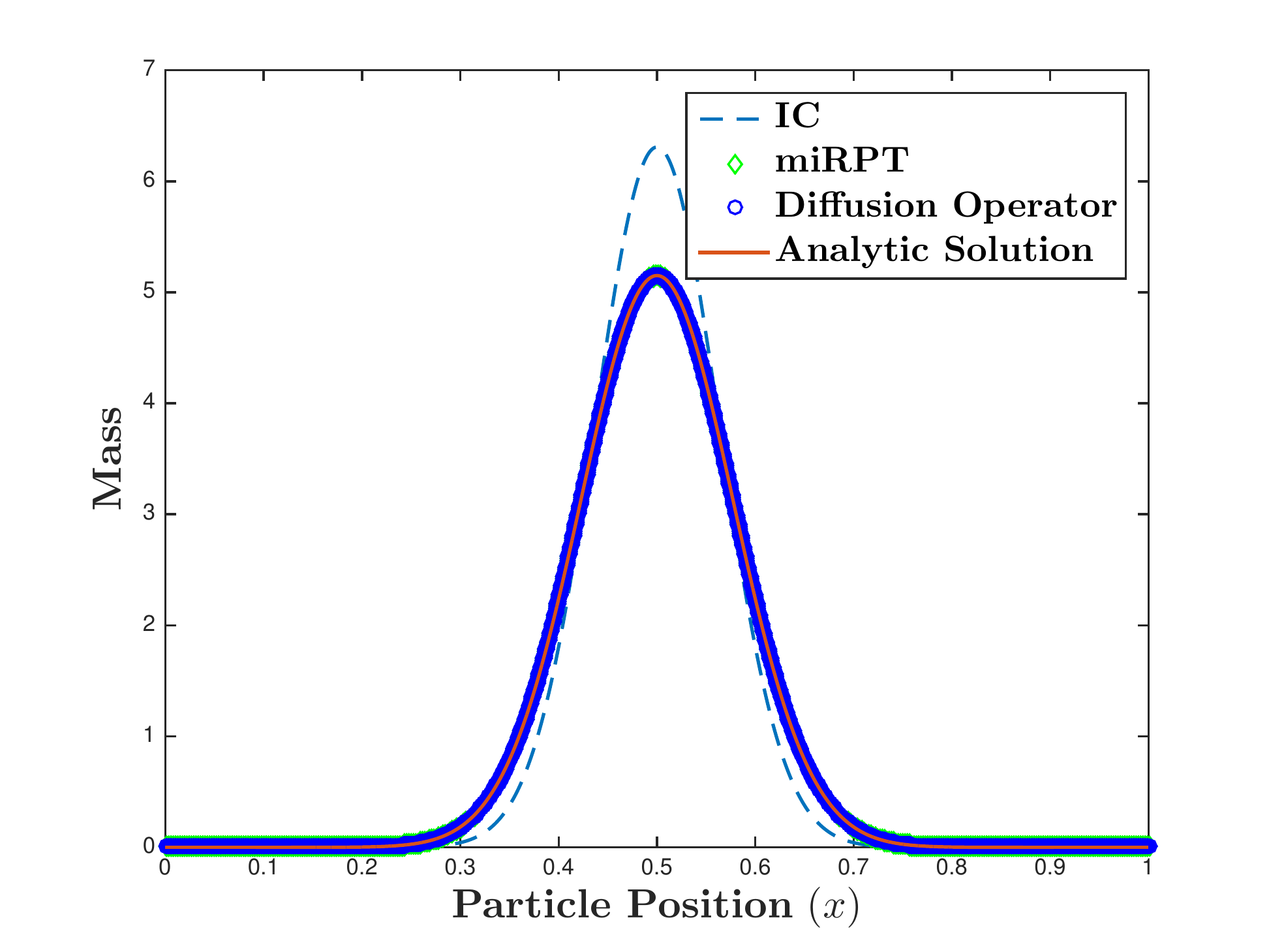} }
    \\
    \subfloat[RMSE vs. $N_M$ (left axis) and stability condition (right axis).]{\includegraphics[width=0.7\textwidth]{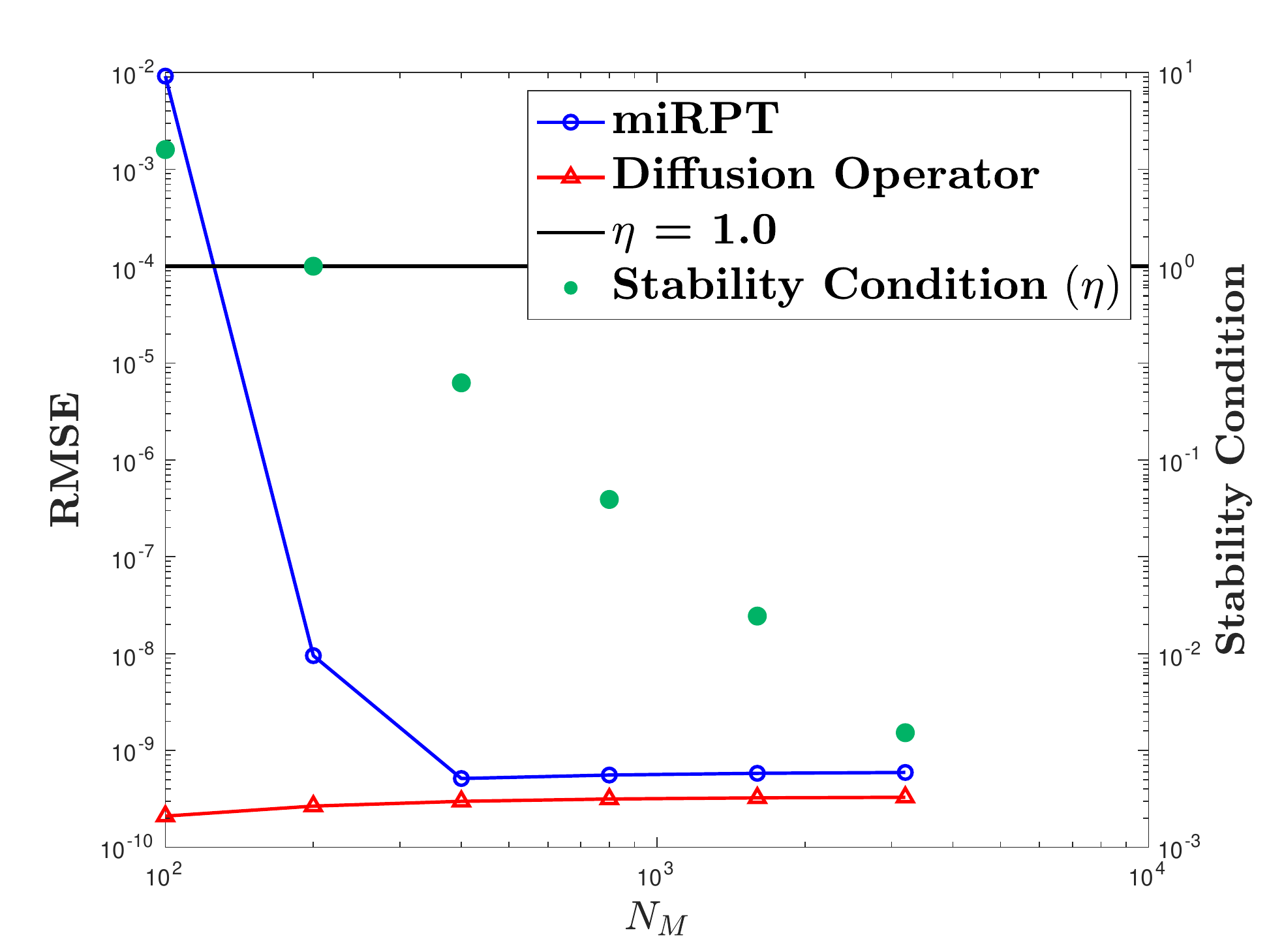} }
    \caption{Error/stability analysis as a function of $N_M$ ($N_I = N_M / 2$) for Gaussian initial condition.}
    \label{fig:equi_Nm_wGauss}
\end{figure}

\begin{figure}[tp]%
    \centering
    \subfloat[Initial condition and final analytic and simulated solutions (plot shows simulated solutions for final data point, $N_M = 3200 = 2 N_I$).]{\includegraphics[width=0.7\textwidth]{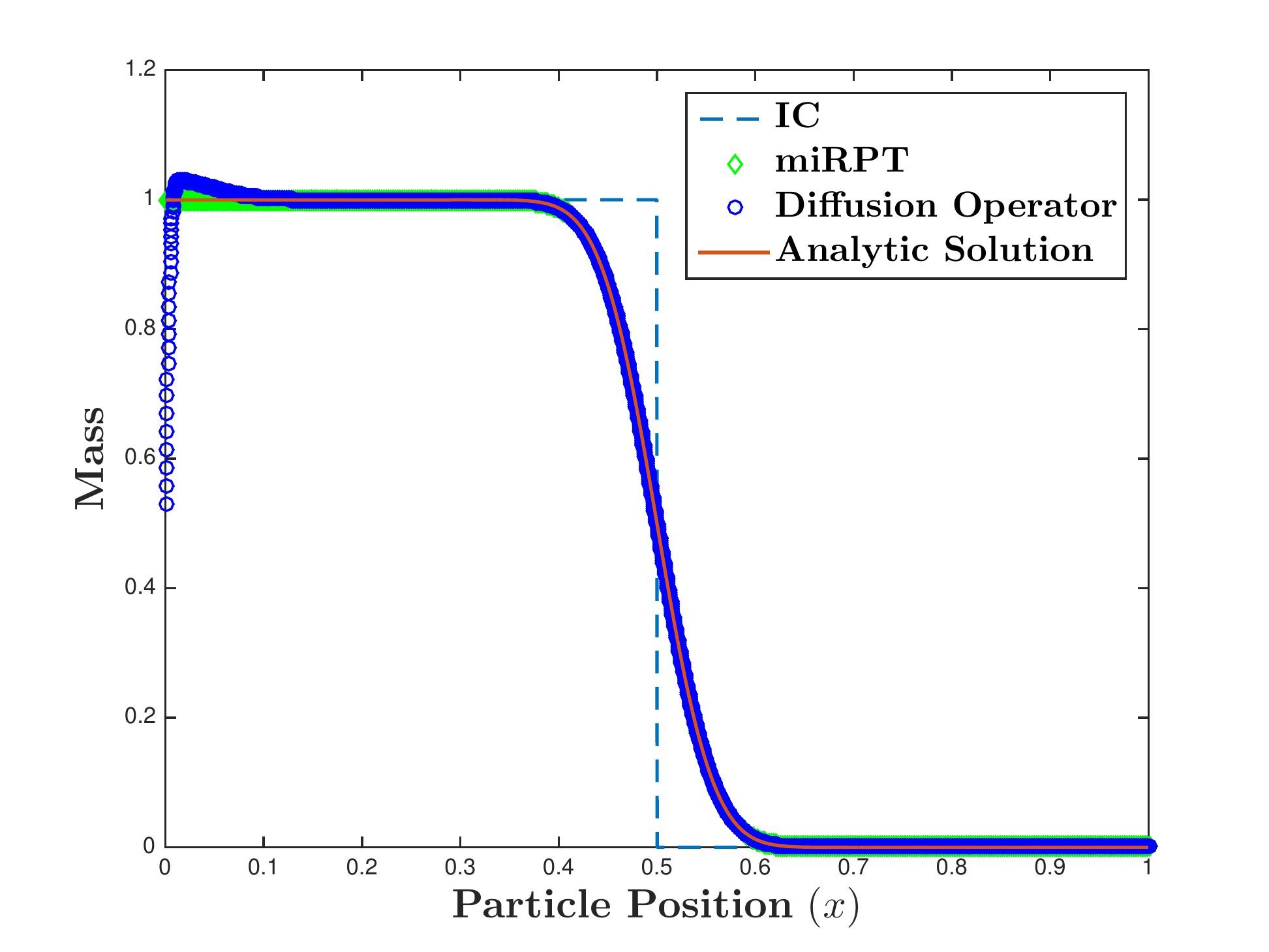} }
    \\
    \subfloat[RMSE vs. $N_M$.]{\includegraphics[width=0.7\textwidth]{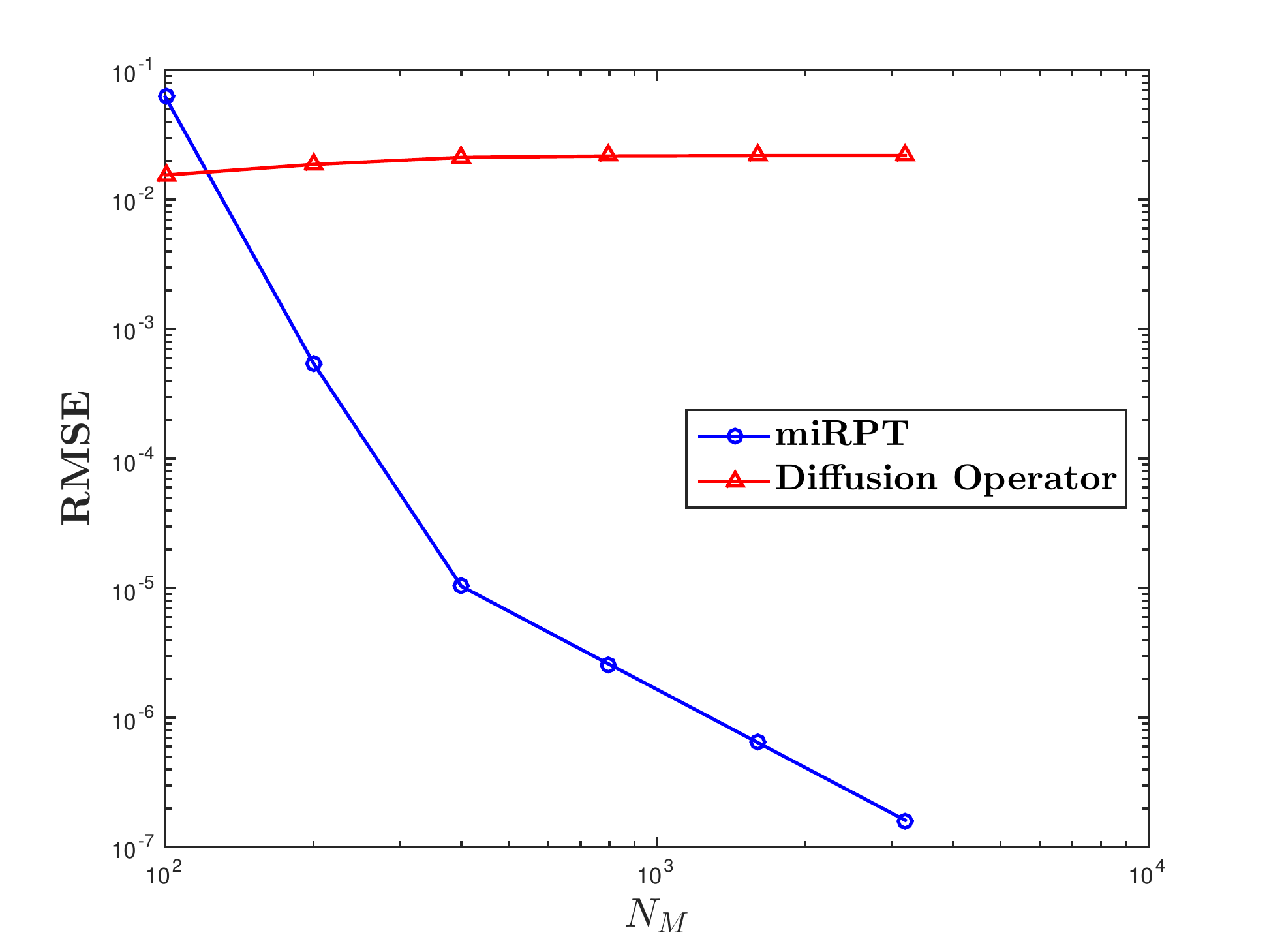} }
    \caption{Error analysis as a function of $N_M$ ($N_I = N_M / 2$) for Heaviside initial condition.}
    \label{fig:equi_Nm_heavi}
\end{figure}

Figures \ref{fig:equi_Nm_wGauss} and \ref{fig:equi_Nm_heavi} show the results of examining the sensitivity of error to the spatial discretization parameters, $N_M$ and $N_I$, the number of mobile and immobile particles, respectively.
For these simulations, $\dt = 1.0 \times 10^{-1}$, $N_I = N_M / 2$ is held at a constant ratio (the reason for this choice of ratio is discussed in the previous section), and $N_M$ is successively doubled.
In Figure \ref{fig:equi_Nm_wGauss} (Gaussian IC) we see similar, but more dramatic, results as in the previous section.
As the choice of $N_M$ drives the miRPT MT algorithm into the stable region of the parameter space (this stability condition is discussed in a subsequent section), we see the error in the miRPT solution decrease by approximately six orders of magnitude and then level off to the minimal value attained by the diffusion operator.
In Figure \ref{fig:equi_Nm_heavi} (Heaviside IC) we see that miRPT MT initially has larger error than the diffusion operator but, the error decreases steadily as particle number increases and becomes many orders of magnitude lower than that achieved by the diffusion operator.

\paragraph{Relationship between error and particle number ratio} 
\label{par:relationship_between_error_and_Nm/Ni}

\begin{figure}[tp]
    \centering
    \includegraphics[width=0.7\textwidth]{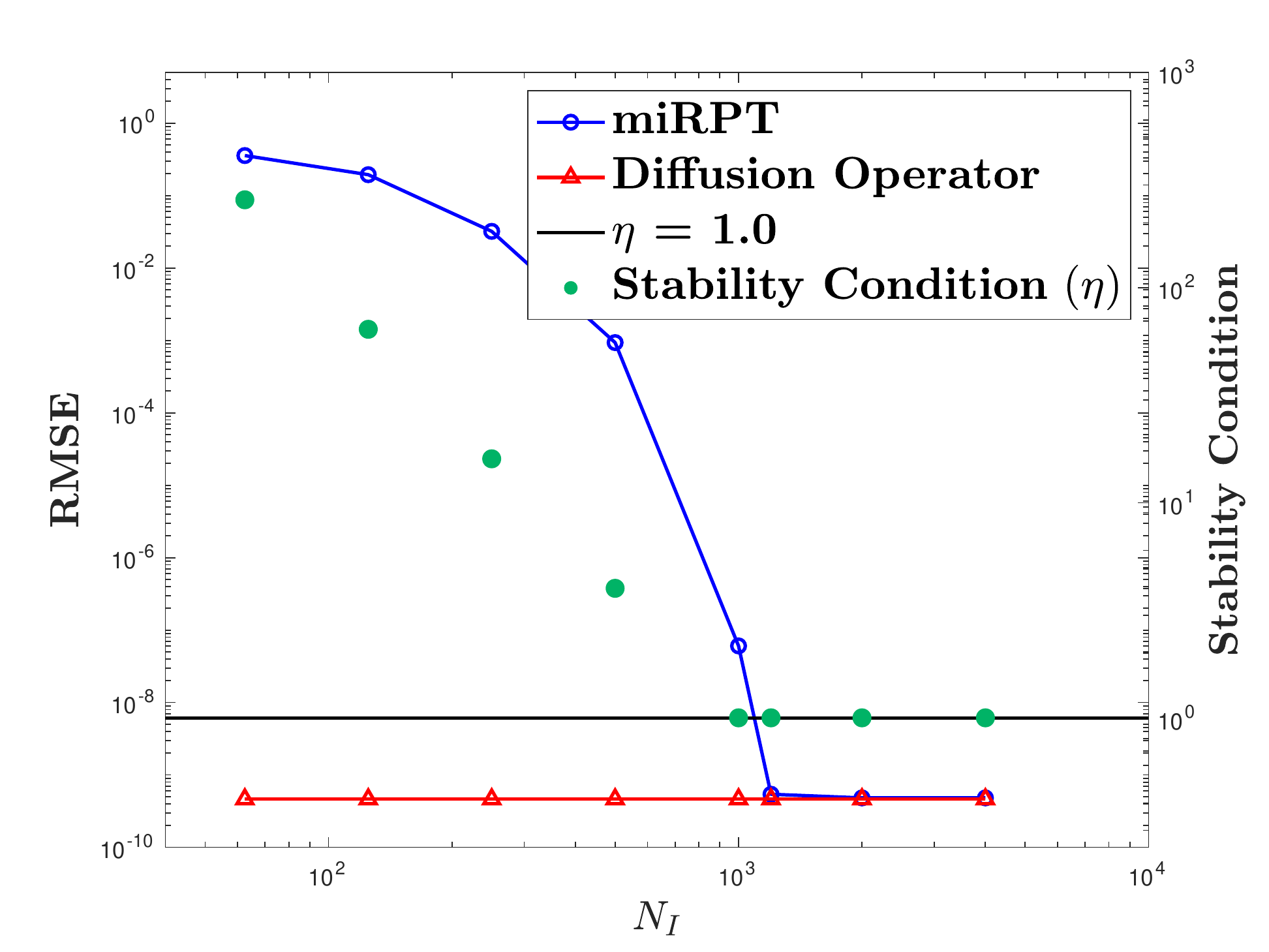}
    \caption{Error/stability analysis as a function of $N_I$ (holding $N_M = 1000$ constant) for Gaussian initial condition with RMSE vs. $N_I$ (left axis) and stability condition (right axis).}
    \label{fig:equi_Ni_wGauss}
\end{figure}

Figure \ref{fig:equi_Ni_wGauss} shows the results of exploring the sensitivity of error to the ratio between the spatial discretization parameters, $N_M$ and $N_I$, the number of mobile and immobile particles, respectively.
For these simulations, only a Gaussian IC is considered, $\dt = 1.0 \times 10^{-3}$, $N_M = 1000$ is held constant, and $N_I$ is successively doubled.
Considering Figure \ref{fig:equi_Ni_wGauss}, we see the error in the miRPT MT algorithm decreasing steadily as $N_I$ increases and the ratio $N_I / N_M$ approaches and exceeds 1.
We also see that the error approaches the minimal level attained by the diffusion operator, which is constant because there are no immobile particles in the diffusion operator approach.
One might have expected the minimal level of error to be attained when $N_I = N_M = 1000$, but for the chosen parameters this is not the case.
To explore this, we added an extra data point $(N_I = 1200)$ to our doubling values of $N_I$, and we see in Figure \ref{fig:equi_Ni_wGauss} that the minimal level of error is achieved relatively soon after we move into the stable region of the parameter space and the ratio $N_I / N_M$ exceeds 1.


\paragraph{Stability condition} 
\label{par:stability_condition}

We see in Figures \ref{fig:equi_dt_wGauss}(b), \ref{fig:equi_Nm_wGauss}(b) and \ref{fig:equi_Ni_wGauss} convincing evidence of something like a stability condition at work in the miRPT MT algorithm.
We formulate this stability condition (in 1D) in the form of an empirical bound on the dimensionless number
\begin{equation}\label{eta1}
    \eta \defeq \frac{(L/\min(N_I, N_M))^2}{D \Delta t} \leq 1,
\end{equation}
where $L$ is the length of the domain, $\Omega$.
The form of $\eta$ is reminiscent of a corresponding von Neumann stability condition, as applied to FD schemes (in fact, it is the reciprocal), where the term $L/\min(N_I, N_M)$ takes the place of $\Delta x$ and can be thought of as the minimum spatial resolution provided by the miRPT MT algorithm (or, alternatively, as the maximum average inter-particle spacing).
We acknowledge that the stability condition proposed here does not operate in exactly the same manner as a von Neumann stability condition for FD schemes.
However, there is numerical intuition behind why this MT stability condition operates, and it lies in the form of the weighting function \eqref{co_loc_density} used for the miRPT mass transfers (i.e., the co-location probability density for a mobile-immobile particle pair).
The argument of the exponential in \eqref{co_loc_density} contains the quantity $-s^2 / (D \dt)$, where $s$ is the distance between a given particle pair.
So, the ratio of squared average inter-particle spacing $\left(L/\min(N_I, N_M)^2\right)$ must stay proportional to $D \dt$.
Otherwise, as this ratio ($\eta$) becomes much larger than 1, the magnitude of mass transfers (or probability of particle interactions) rapidly becomes, on average, indistinguishable from 0, to machine precision.

We see, in all of the aforementioned figures, a distinct drop in the error (sometimes several orders of magnitude) when $\eta$ crosses below the threshold value of 1, where, afterward, the error appears to converge to a minimal level.
As well, we note here that for the results depicted in Figure \ref{fig:equi_dt_heavi}(b), there is in fact a distinct, relative jump in error for the final data point, where $\eta > 1$; however, the scaling of the figure does not make this apparent.
The results shown in Figure \ref{fig:equi_Ni_wGauss} inform us that there may be something slightly more complex occurring in regards to the stability condition.
In Figure \ref{fig:equi_Ni_wGauss}, we do see a marked drop in error (roughly 4 orders of magnitude) between data points 4 and 5 ($N_I = 500, 1000$, respectively), when the stable region of parameter space is entered.
However, the minimal level of error is not attained until the value of $N_I$ is increased to 1200.
The form of $\eta$ also provides insight as to why the error in the miRPT MT algorithm is relatively insensitive to changes in $\dt$ yet highly sensitive to varying particle number (whether mobile or immobile) because the stability condition varies linearly with $\dt$ and quadratically with the minimum particle discretization.

\subparagraph{Stability in 2D} 
\label{subp:stability_in_2d}

\begin{figure}[tp]
    \centering
    \includegraphics[width=0.7\textwidth]{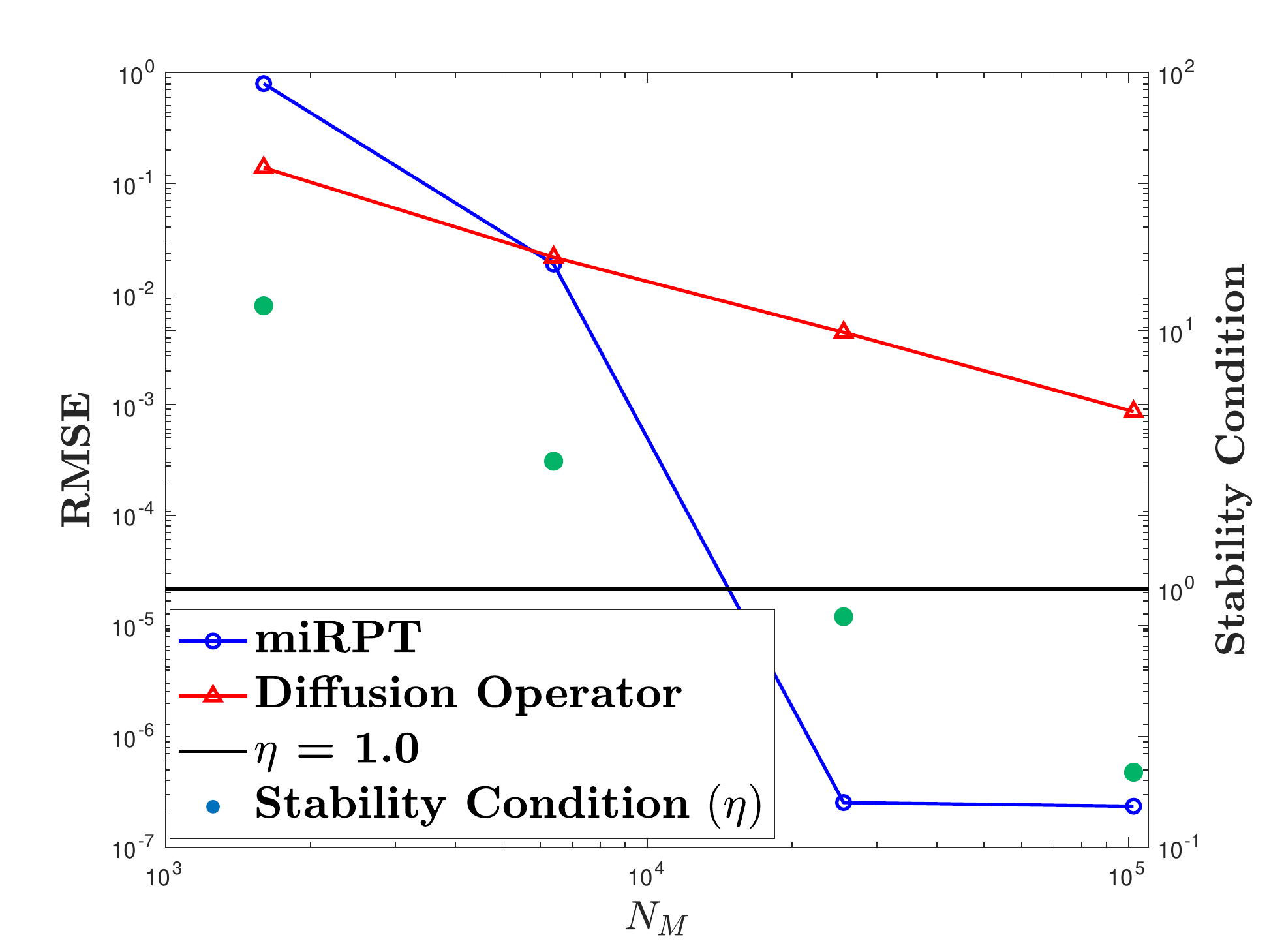}
    \caption{Error/stability analysis as a function of $N_M$ for 2D Gaussian initial condition with RMSE vs. $N_M$ (left axis) and stability condition (right axis).}
    \label{fig:2D_stability}
\end{figure}

We see the results of the stability analysis for a 2D simulation depicted in Figure \ref{fig:2D_stability}, where the form of the stability condition becomes
\begin{equation}\label{eta2}
    \eta \defeq \frac{(\max(L_1, L_2)/\min(\sqrt{N_I}, \sqrt{N_M}))^2}{D \Delta t} \leq 1,
\end{equation}
where $L_1$ and $L_2$ are the lengths of the sides of the rectangular 2D domain, $\Omega$.
The changes in the numerator of \eqref{eta2}, as compared to \eqref{eta1}, serve to, again, allow this quantity to represent the minimum spatial resolution provided by the simulation parameters (corresponding to $\Delta x$ in an Eulerian framework).
For the results shown in Figure \ref{fig:2D_stability}, $L_1 = L_2 = 1$, , $\dt = 1.0 \times 10^{-1}$, $N_I = N_M / 2$, and $\sqrt{N_M}$ was successively doubled.
We see in Figure \ref{fig:2D_stability}, that for the properly formulated stability condition, there is the anticipated drop in error ($\approx$ 5 orders of magnitude) when crossing from the unstable to stable region of parameter space.
However, since the form of $\eta$ is changed for the 2D case, we no longer have a quadratic decrease in $\eta$ with particle number but only a linear relation.
As a result, driving the algorithm into the stable region requires much larger increases in particle number than it does in the 1D case.

\subsubsection{Mass-transfer analysis for randomly-spaced particles} 
\label{ssub:mass_transfer_analysis_for_randomly_spaced_particles}

\begin{figure}[tp]%
    \centering
    \subfloat[Initial condition and final analytic and simulated solutions (plot shows simulated solutions for final data point, $N_M = 12800 = N_I$).]{\includegraphics[width=0.7\textwidth]{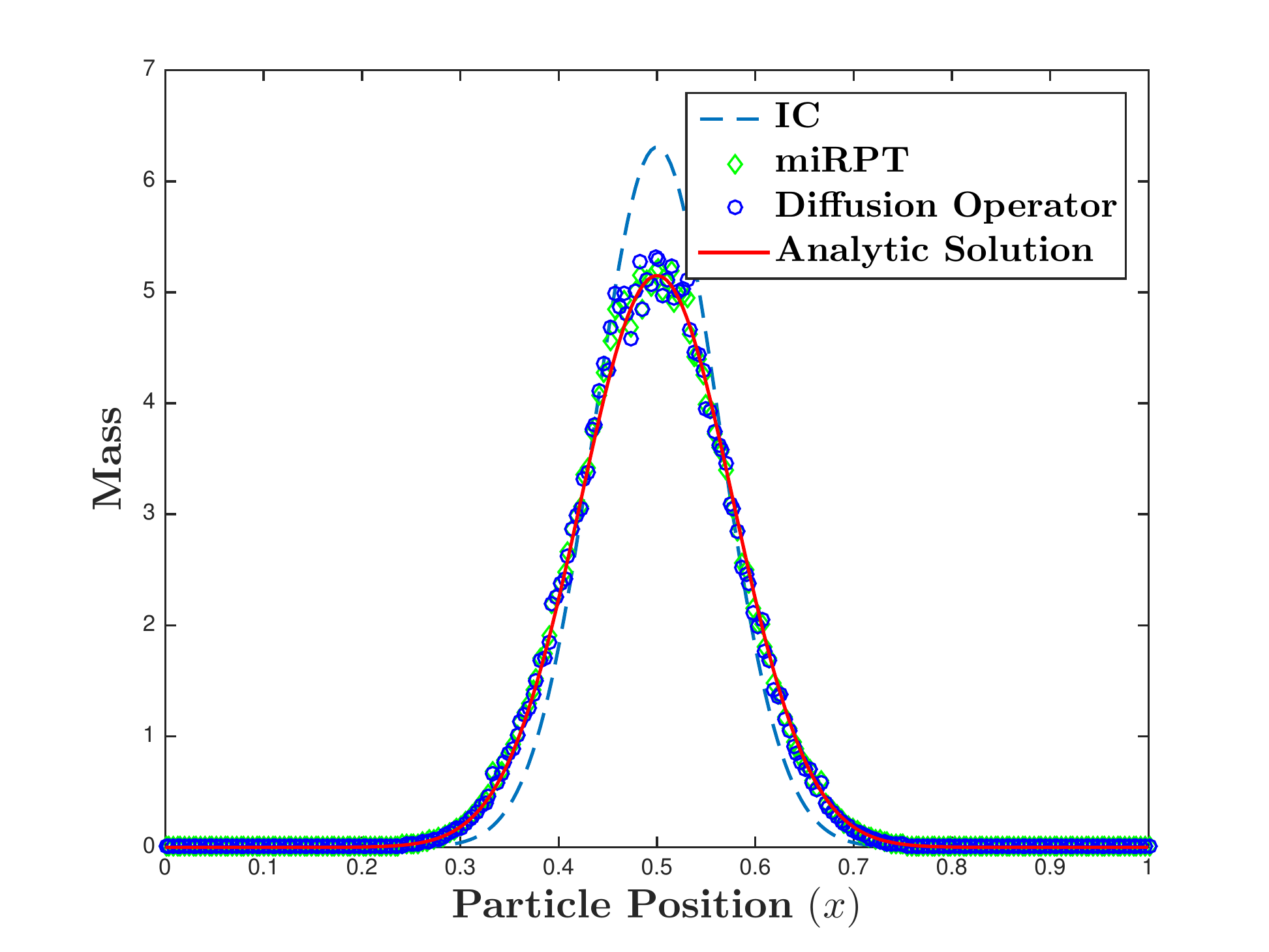} }
    \hspace{0.01\textwidth}
    \subfloat[RMSE vs. $N_M$ (left axis) and stability condition (right axis).]{\includegraphics[width=0.7\textwidth]{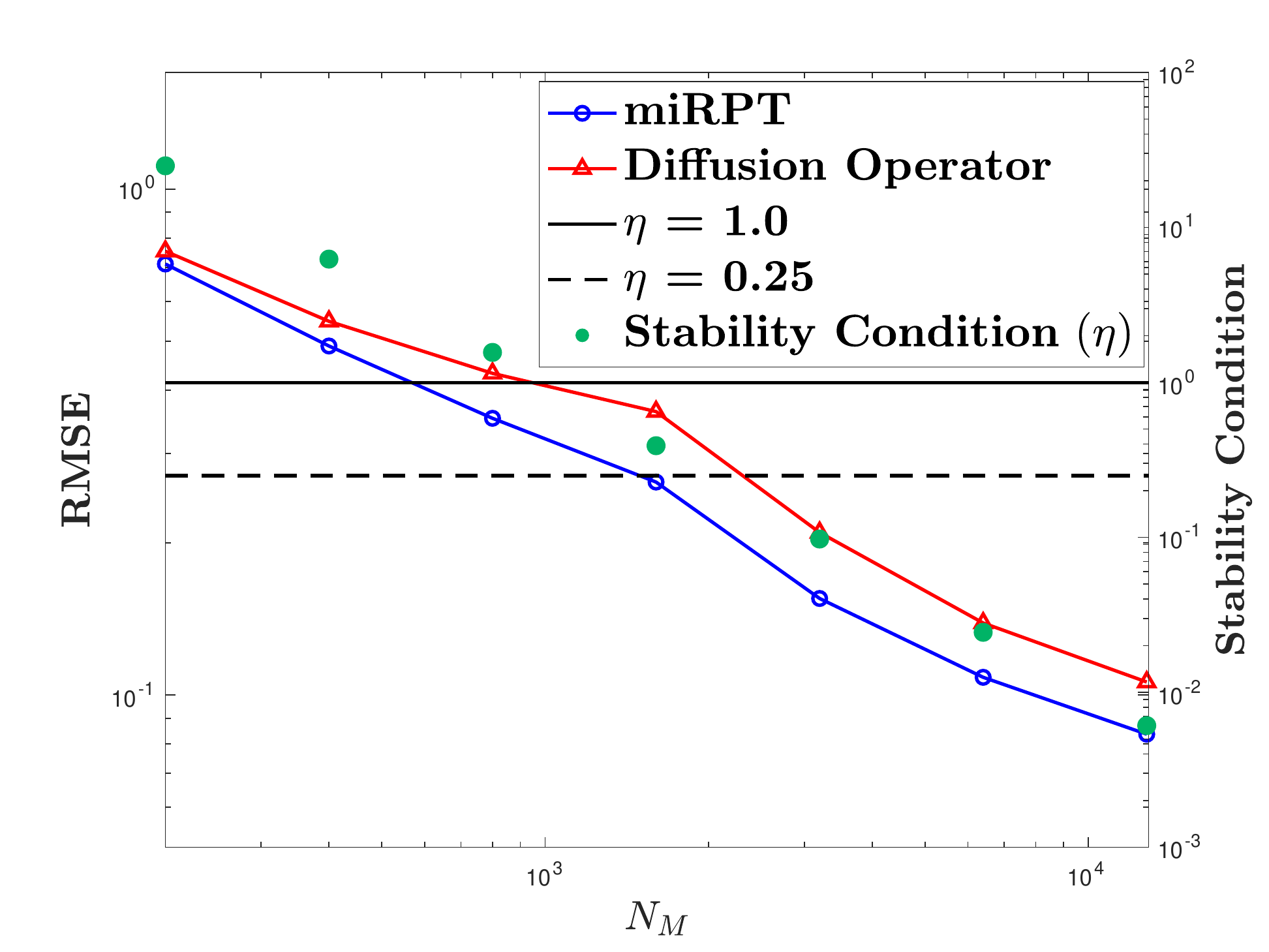} }
    \caption{Error/stability analysis as a function of $N_M$ for Gaussian initial condition using randomly-spaced particle positions. Results shown are an ensemble average of 100 simulations per value of $N_M$, and particles were ``binned'' into 250 equally-spaced bins to average out random variations.}
    \label{fig:ens_run_gaussian}
\end{figure}

\begin{figure}[tp]%
    \centering
    \subfloat[Initial condition and final analytic and simulated solutions.]{\includegraphics[width=0.7\textwidth]{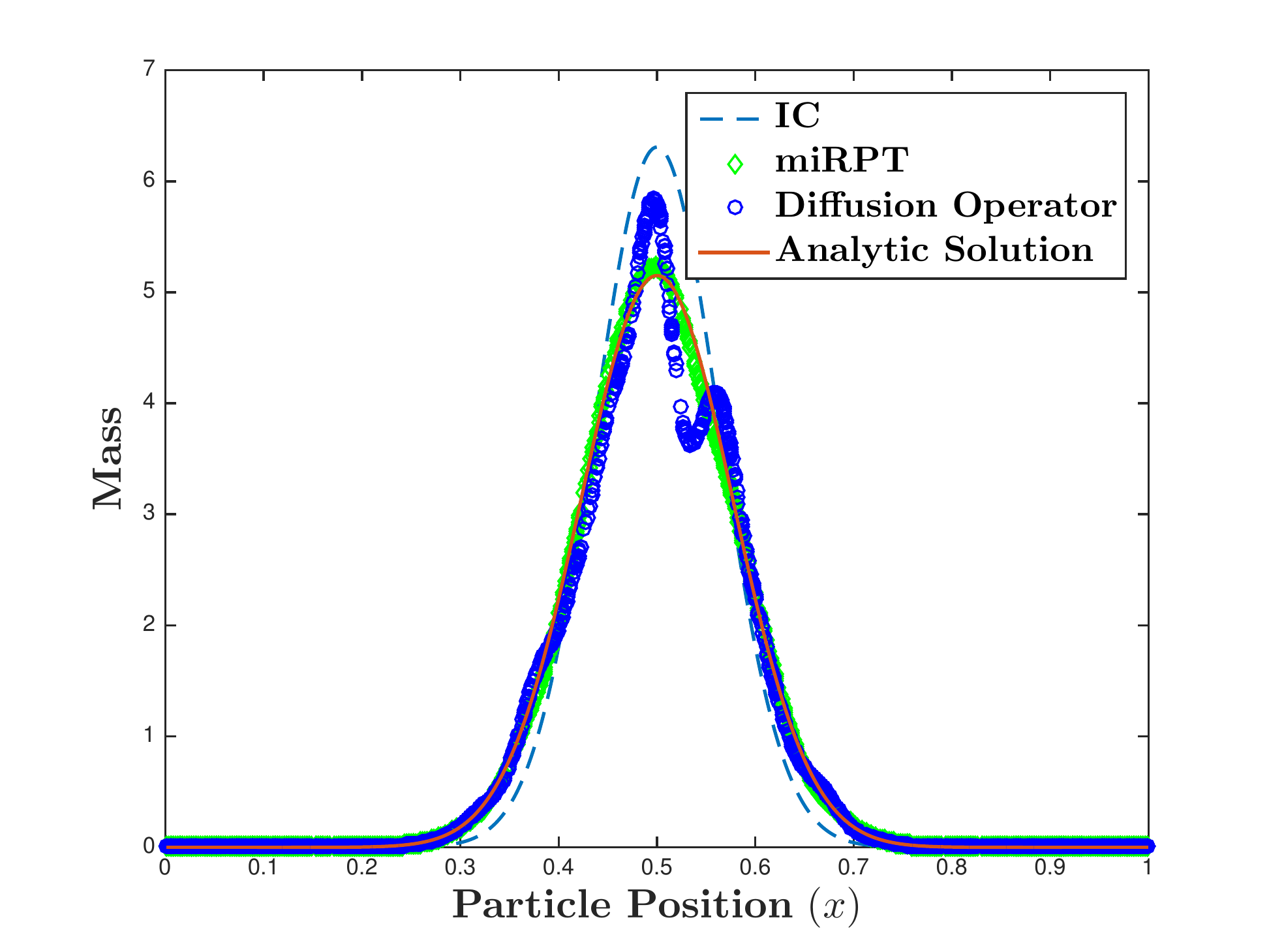}}
    \hspace{0.01\textwidth}
    \subfloat[Spatial variance increase vs. time.]{\includegraphics[width=0.7\textwidth]{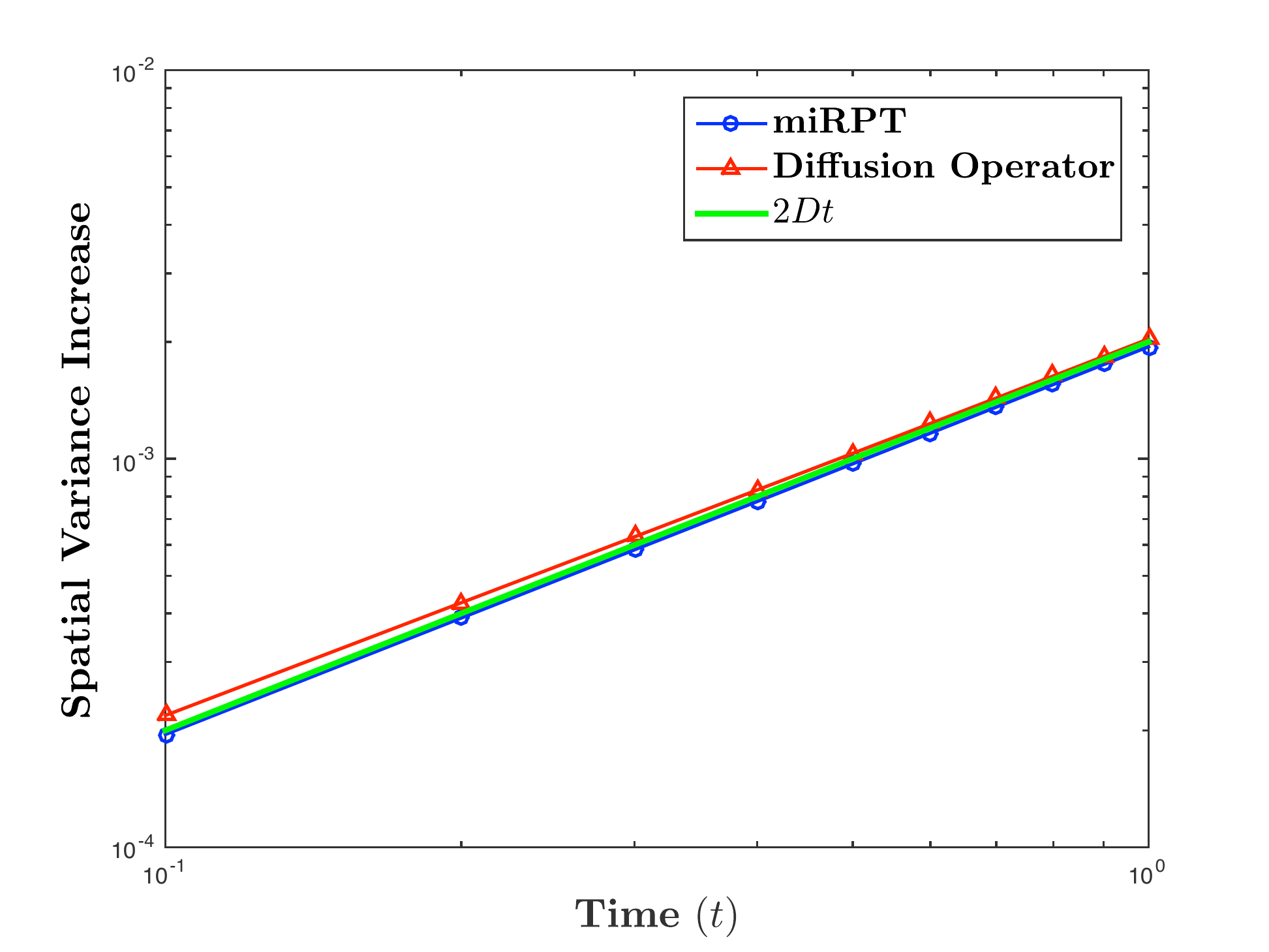} }
    \caption{Error analysis in terms of variance increase with $N_I = N_M = 1600$ for Gaussian initial condition.}
    \label{fig:var_gaussian}
\end{figure}

\begin{figure}[tp]%
    \centering
    \subfloat[Initial condition and final simulated solutions.]{\includegraphics[width=0.7\textwidth]{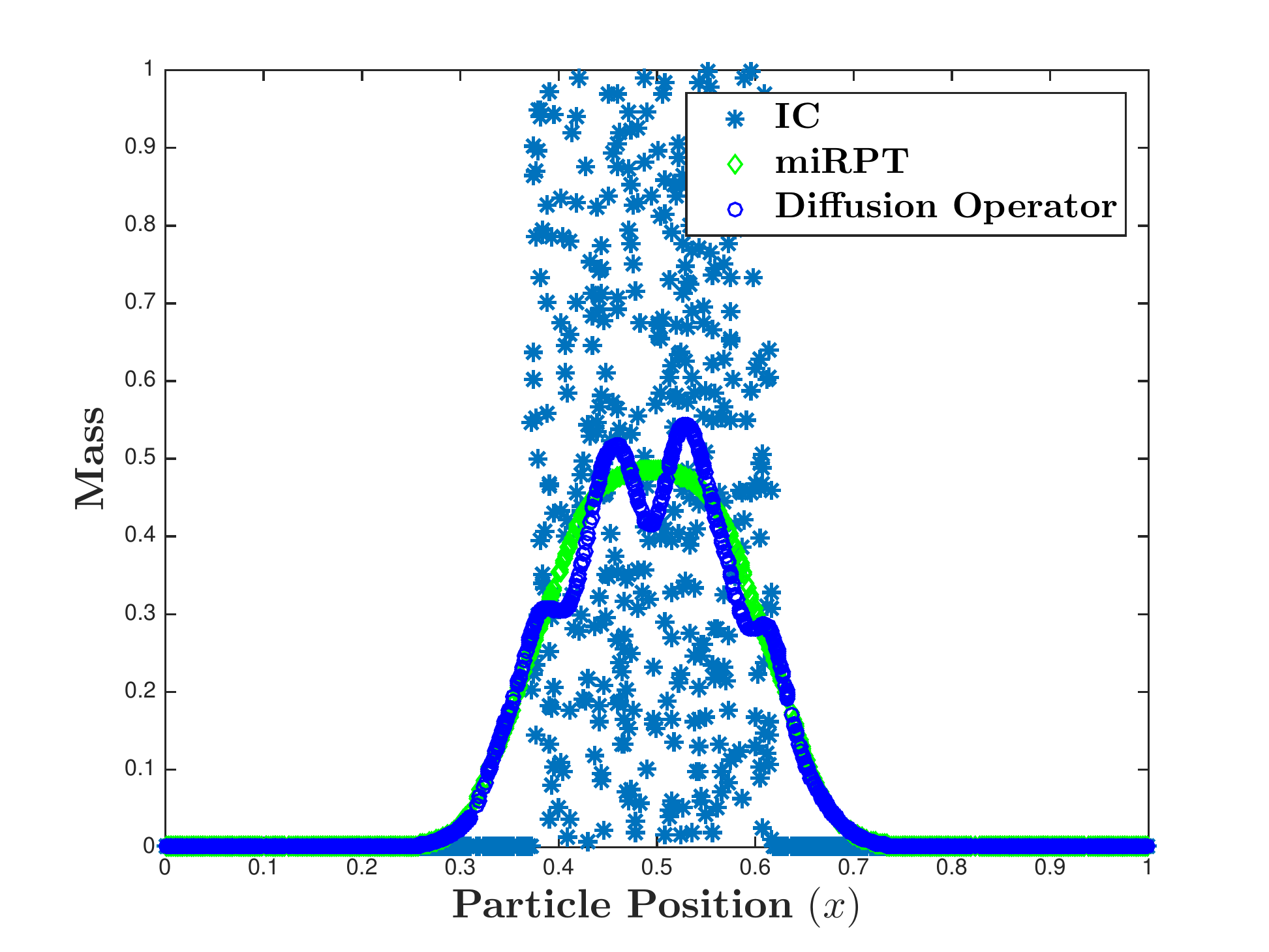}}
    \hspace{0.01\textwidth}
    \subfloat[Spatial variance increase vs. time.]{\includegraphics[width=0.7\textwidth]{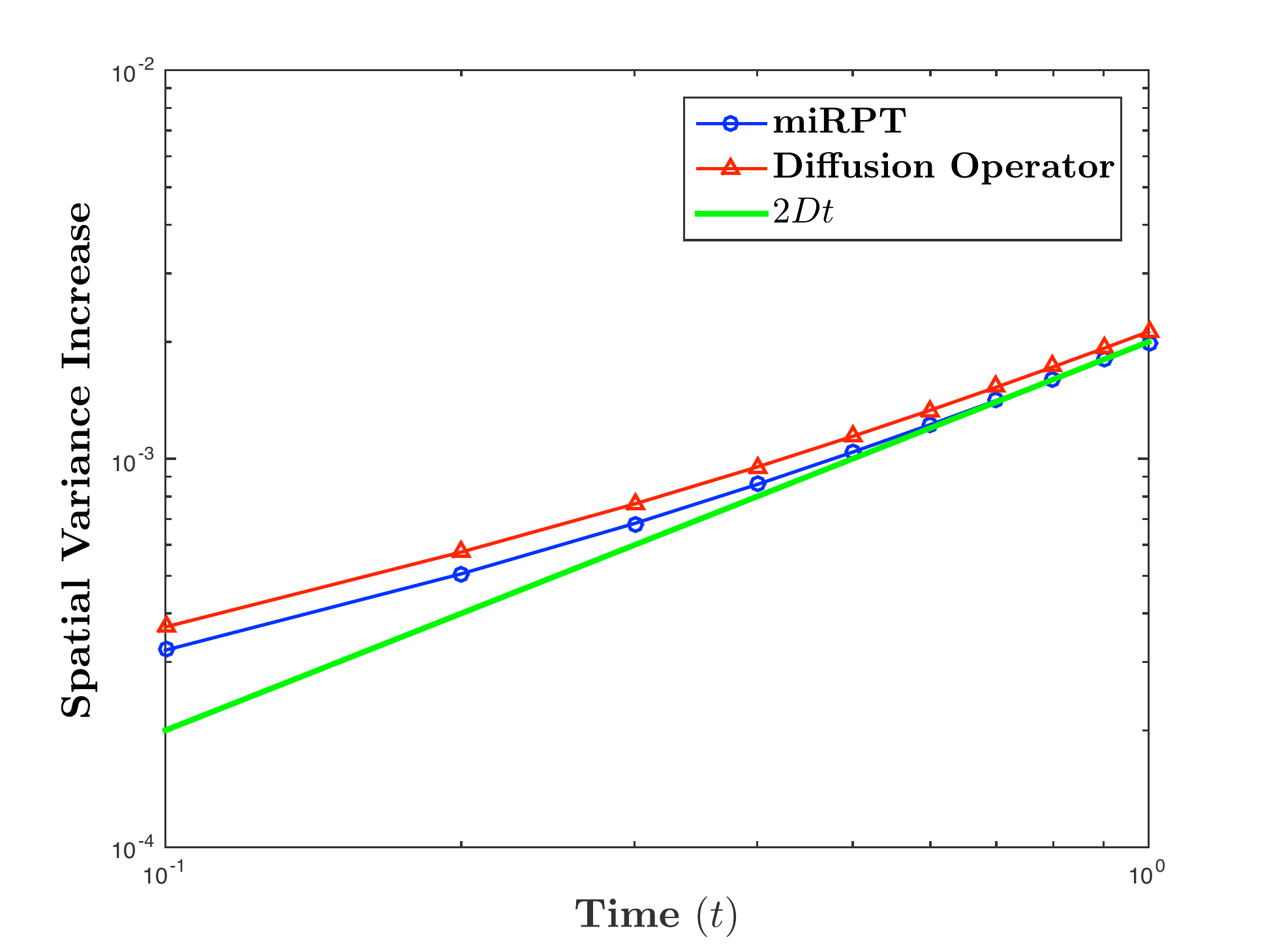}}
    \caption{Error analysis in terms of variance increase with $N_I = N_M = 1600$ and $\dt = 1.0 \times 10^{-1}$ for noisy box initial condition.}
    \label{fig:var_noisy_box1}
\end{figure}

\begin{figure}[tp]%
    \centering
    \subfloat[Initial condition and final simulated solutions.]{\includegraphics[width=0.7\textwidth]{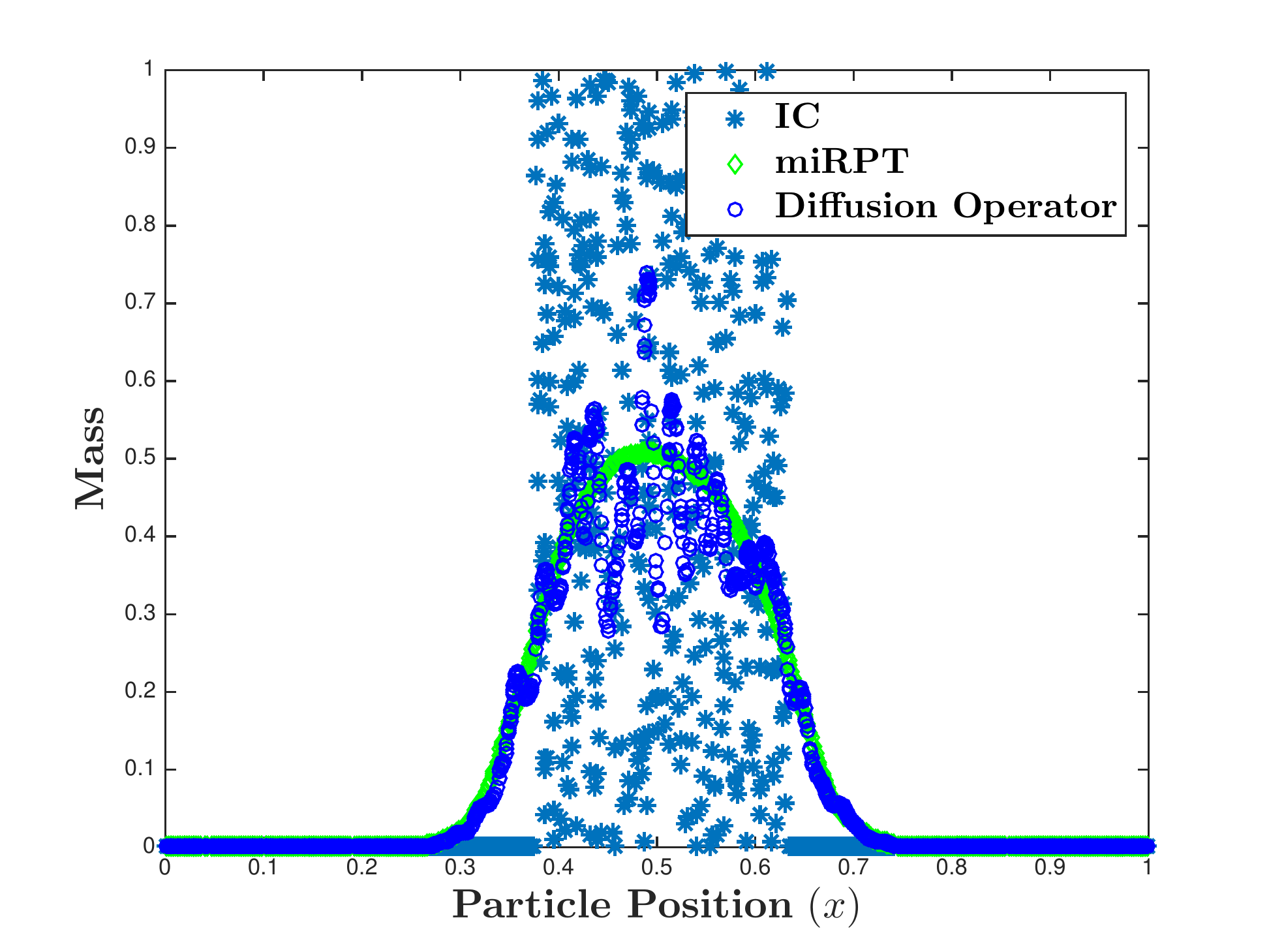}}
    \hspace{0.01\textwidth}
    \subfloat[Spatial variance increase vs. time.]{\includegraphics[width=0.7\textwidth]{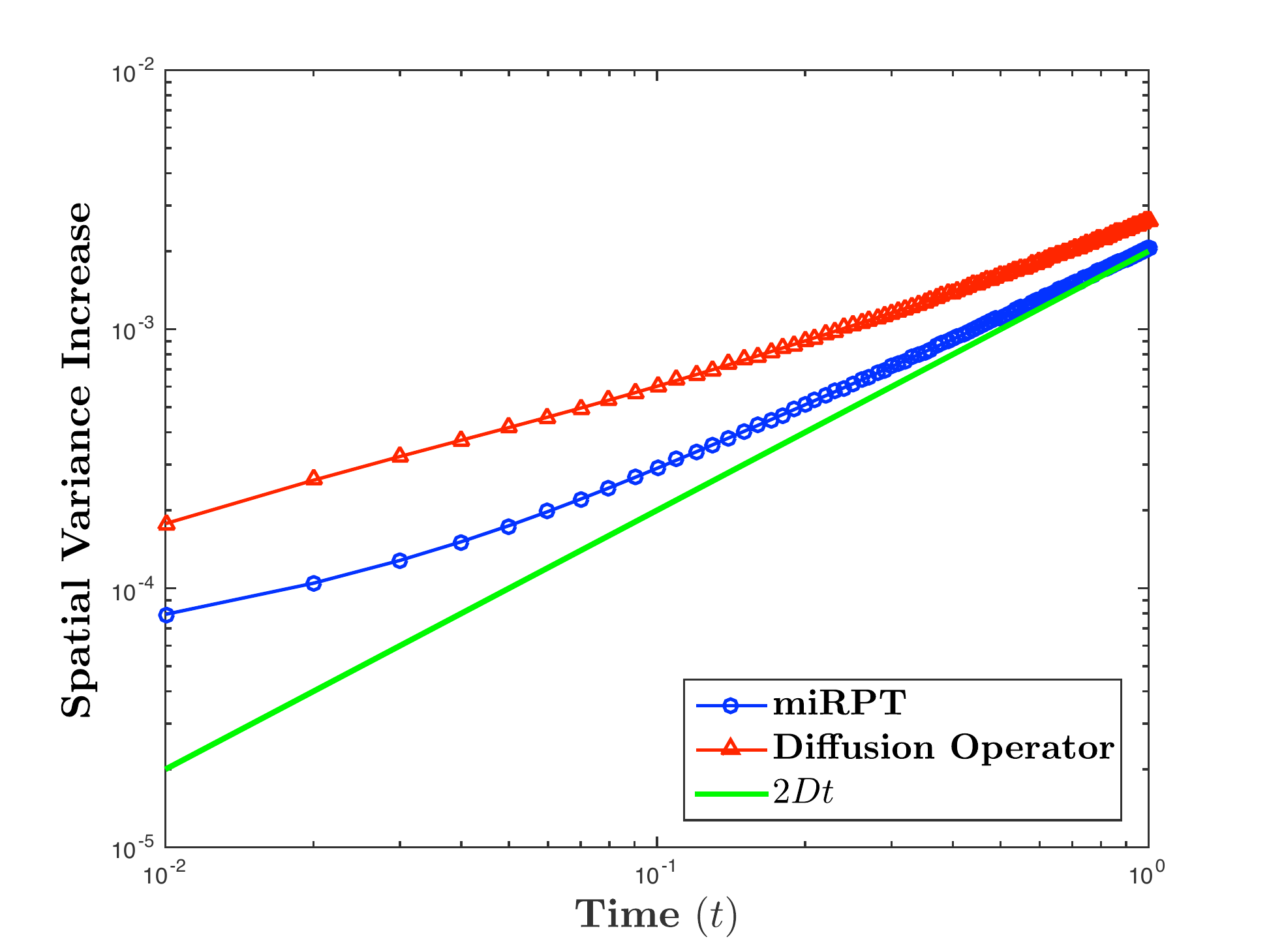} }
    \caption{Error analysis in terms of variance increase with $N_I = N_M = 1600$ and $\dt = 1.0 \times 10^{-2}$ for noisy box initial condition.}
    \label{fig:var_noisy_box2}
\end{figure}

Having verified that the miRPT MT algorithm solves the diffusion equation satisfactorily in the idealized, equally-spaced particle case, we must also consider the case that corresponds to mobile particles moving via random walk.
For this analysis, immobile particles are still spaced evenly across the domain, but the mobile particles are assigned positions according to independent draws from a Uniform, $\cU(0, 1)$, distribution to \revtwo{recreate the irregular spatial distribution that would be induced by random walks.}
We perform a convergence analysis for the randomly-spaced particles in two cases.
First, we conduct an ensemble run of 100 simulations for each set of parameters and average the results to see whether the expectation of these stochastically perturbed simulations converges to the expected value of the analytic solution (Figure \ref{fig:ens_run_gaussian}).
Next, we examine the results of a single randomly-spaced particle simulation from a statistical standpoint to confirm that the spatial variance in mass increases at the expected rate of $2 D t$.
This second analysis is conducted both for a Gaussian IC (Figure \ref{fig:var_gaussian}) with an analytic solution that can be visually verified and also for a non-analytic ``noisy box'' IC (Figures \ref{fig:var_noisy_box1} and \ref{fig:var_noisy_box2}).

\paragraph{Ensemble run for Gaussian initial condition} 
\label{par:ensemble_run_for_gaussian_ic}

The results of the 100-member ensemble run for a Gaussian IC are shown in Figure \ref{fig:ens_run_gaussian}.
For this simulation, $\dt = 1.0 \times 10^{-3}$, $N_I = N_M$, and $N_M$ was successively doubled.
The results of this ensemble run were then averaged by first placing particles into 250 spatial ``bins'' and averaging the masses in each bin across the ensemble to generate an ensemble-averaged concentration for each bin.
This binning was necessary because, given randomly-assigned particle positions from a continuous probability distribution, a certain particle will never be in the same position for any two simulations, so a particle-wise average does not make sense.

While strict convergence is difficult to prove for a stochastic algorithm, we do see favorable results in Figure \ref{fig:ens_run_gaussian}.
First, in Figure \ref{fig:ens_run_gaussian}(a), for the final refinement in particle number ($N_I = N_M = 12800$), we see good visual agreement between the analytic solution and both the miRPT MT algorithm and the diffusion operator approach.
However, as would be expected, the error shown in Figure \ref{fig:ens_run_gaussian}(b) does not drop nearly as quickly with increases in particle number as it did for the equally-spaced case.
Additionally, we see miPRT MT outperforming the diffusion operator approach for all tested parameters.
Further, there is evidence to suggest that the stability condition may become more restrictive with randomly-spaced particles.
In Figure \ref{fig:ens_run_gaussian}(b), the largest drop in error is seen between points 4 and 5 ($N_M = 1600$ and $3200$, respectively), both of which are associated with values of $\eta < 1$, suggesting that $\eta < 0.25$ may be better practice when particle positions are not equally-spaced.
This follows logically, since simulations with randomly-spaced particles will inevitably include some larger gaps between particles (\revtwo{which, some argue, physically represents a poorly mixed area of the domain}).
As a result, the quantity $L/\min(N_I, N_M)$ (representing maximum \emph{average} particle spacing) must be made smaller than in the equally-spaced case.

\paragraph{Variance analysis} 
\label{par:variance_analysis}

The results of a variance analysis for a single run with randomly-spaced mobile particles are shown in Figures \ref{fig:var_gaussian}-\ref{fig:var_noisy_box2}.
The spatial variance in mass is calculated as
\begin{equation}\label{spatial_variance}
    \sigma^2_S = \sum_{i = 1}^{N_I} \frac{m_i^{(I)}}{m_0^{(I)}} \left(x_i^{(I)} - \bar m \right)^2,
\end{equation}
where $m_0^{(I)} \defeq \sum_{i = 1}^{N_M} m_i^{(I)}$ and $\bar m$ is the first spatial moment, or center of mass.
Here we note that the center of mass should be 0.5 for all results in this section; however, we do not plot results for mean position, as spatial mean was nearly exact for the miRPT MT algorithm and gained a minimal amount of error with time for the diffusion operator approach.
These results may be visually inferred by the shape of the final mass distributions in Figures \ref{fig:var_gaussian}(a)-\ref{fig:var_noisy_box2}(a).
For ease in visual comparison, we depict spatial variance \emph{increase} in Figures \ref{fig:var_gaussian}-\ref{fig:var_noisy_box2} that is computed by subtracting the spatial variance of the IC from $\sigma^2_S$.
In addition, the single-realization results, as shown in this section, will vary based on initial mobile particle positions, but the results depicted in the figures are typical for the given parameters and ICs.

For these simulations, $N_I = N_M = 1600$, $\dt = 1.0 \times 10^{-1}$ for Figures \ref{fig:var_gaussian} and \ref{fig:var_noisy_box1}, and $\dt = 1.0 \times 10^{-2}$ for Figure \ref{fig:var_noisy_box2}.
In Figure \ref{fig:var_gaussian}(b), we see that for the Gaussian IC, the miRPT MT algorithm is virtually error-free in terms of variance increase and that a small amount of error is introduced by the diffusion operator approach.
In Figures \ref{fig:var_noisy_box1} and \ref{fig:var_noisy_box2} we consider a non-analytic ``noisy box'' IC, wherein the middle 20\% of particles are assigned their initial mass according to independent draws from a $\cU(0, 1)$ distribution.
This is done so as to rule out finite boundary effects while eliminating the smooth gradients provided by the Gaussian IC.
Considering Figure \ref{fig:var_noisy_box1}, we see that both miRPT MT and the diffusion operator initially introduce extra spatial variance; although, since this error stays relatively constant over the 1 second simulation, the error actually \emph{decreases}, in a relative sense, with time.
A somewhat different result can be seen in Figure \ref{fig:var_noisy_box2}, where the diffusion operator at first increases the spatial variance by nearly \revtwo{an order of magnitude}, and miRPT MT to a lesser degree.
Both algorithms, however, closely approach the analytic variance after a certain point in time, with miRPT MT becoming nearly error-free by the end of the simulation.
This notion of ``accuracy as a function of repeated operation'' is discussed in \cite{mass_trans_acc}.

\subsection{Reactive transport model} 
\label{sub:reactive_transport_model}

Having established that the miRPT MT algorithm is capable of solving the diffusion equation with a sufficient level of accuracy, we now wish to apply the full miRPT algorithm (all of the steps depicted in Figure \ref{fig:imRPT_chart}) to a chemically-complex reactive transport problem.
The test problem we choose is based on one considered by \emph{Leal et al.} \cite{Leal_2016_xLMA}.
In this system, which we refer to as the calcite-dolomite reactive transport (CDRT) system, we consider a high temperature and pressure ($60\degree$C, 100 bar) 1-dimensional domain (0.5 m in length) composed of calcite (CaCO$_3$) and quartz (SiO$_2$) (2\% and 98\%, respectively, at uniform 50\% porosity) into which a constant concentration of CO$_2$-saturated brine composed of various salts ($\approx$ 0.88 mol/L NaCl, 0.05 mol/L MgCl, 0.01 mol/L CaCl$_2$) is injected at $x = 0$.
Injection of an acidic brine with Mg concentrations higher than in the initial domain causes the fluid to become undersaturated with respect to calcite and oversaturated with respect to dolomite, resulting in the co-located dissolution of calcite and precipitation of dolomite as a reaction front that advances through the domain in the direction of advection in time.
Calcite dissolution buffers the pH at the leading edge of this reaction front, while dolomite dissolution buffers the pH within the reaction front.
Once both calcite and dolomite fully dissolve at any given location, \revtwo{the pH is no longer buffered} and is equal to the pH of the inlet brine ($\approx$ 3.0), defining the trailing edge of the reaction front.
For a comparable schematic diagram of this IC and boundary condition (BC), see Figure 1 in \cite{Leal_2016_xLMA}.
This system is governed by the ADRE given in \eqref{adre}, where $v = 2.4 \times 10^{-5}$ m/s and $D = 1.2 \times 10^{-7}$ m$^2$/s.
Reactions are considered to be instantaneous, and these calculations are handled by an operator-splitting approach using the PhreeqcRM geochemical solver \cite{phreeqcrm} and the \texttt{phreeqc.dat} thermodynamic database \cite{PHREEQC}.

This CDRT system is an ideal test bed for miRPT for a few reasons.
First, calcite and dolomite are solid phases that are not transported but must interact with the flowing CO$_2$ and salts in the injection brine that are being transported by advection and diffusion.
Thus, calcite and dolomite concentrations will always be stored on immobile particles. The masses of the aqueous ions composing the injection brine (10 total species) will be stored on mobile particles for transport (Steps (a)-(b) in Figure \ref{fig:imRPT_chart}) and then transferred to the immobile particles using the miRPT MT algorithm (Step (c) in Figure \ref{fig:imRPT_chart}).
Once there, reaction mass-balance calculations are conducted before the resulting aqueous ions are transferred back, again according to miRPT MT (Steps (d)-(h) in Figure \ref{fig:imRPT_chart}).
Another reason this problem was chosen is that the low level of diffusion $\left(\cO\left(10^{-7}\right) \text{m}^2/\text{s} \right)$ will present a challenge for miRPT MT, due to the stability condition given in \eqref{eta1}.
As discussed in Section \ref{par:ensemble_run_for_gaussian_ic}, when particles are randomly-spaced (as will be induced by the diffusive random walks in this model), the stability condition becomes more restrictive than imposing $\eta < 1$, possibly requiring $\eta < 0.25$.
For that reason, we devote \ref{sub:selection_of_model_parameters} to choosing the proper model parameters to ensure a stable simulation.

\begin{sidewaysfigure}[tp]%
    \includegraphics[width=\textwidth]{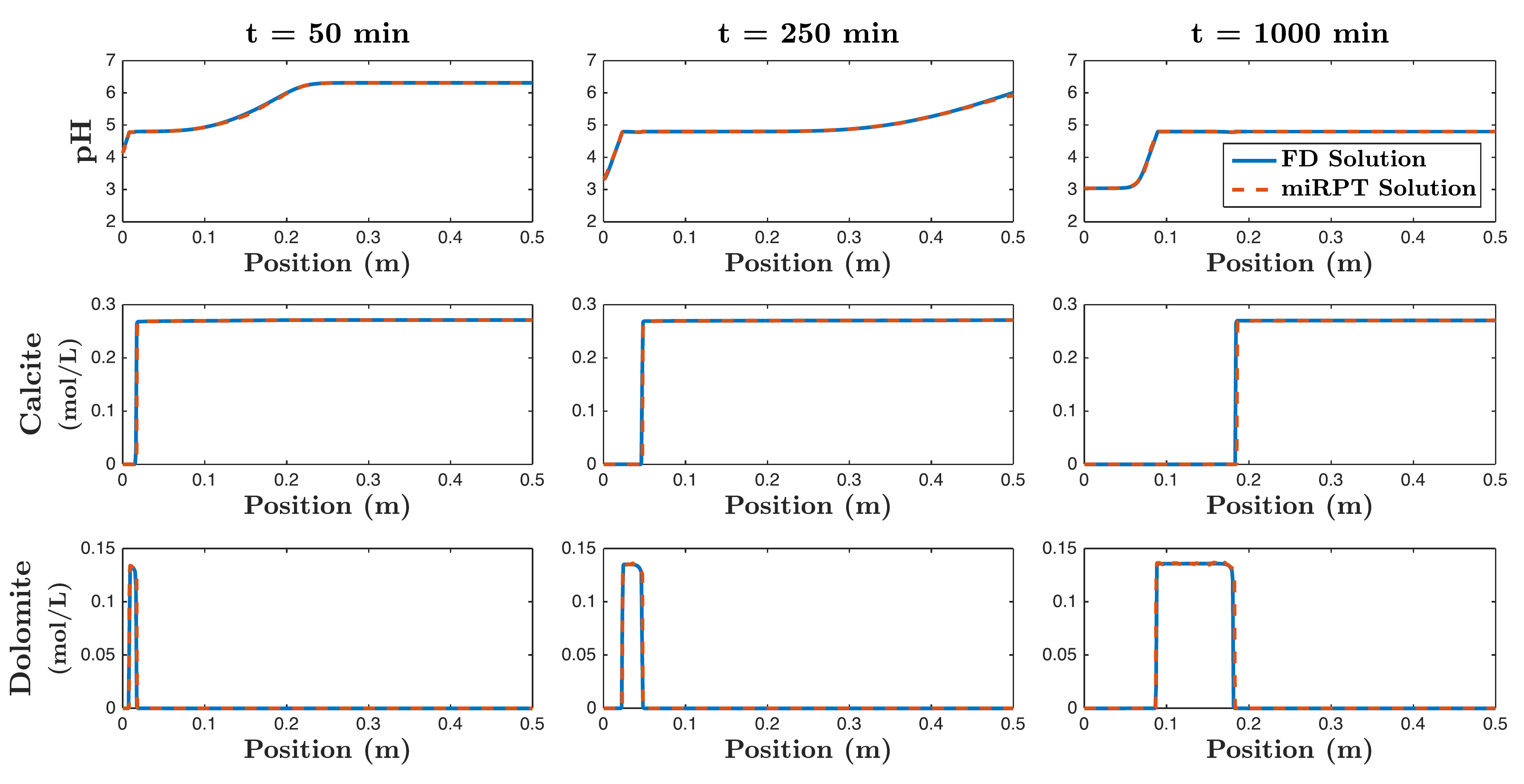}
    \caption{Results for calcite-dolomite reactive transport system, comparing FD and miRPT results. pH, calcite concentration, and dolomite concentration in 0.5 m domain are depicted for 50, 250, and 1000 minutes of simulation time. Acidity moving left to right (top row) dissolves calcite (middle), immediately precipitating dolomite (bottom) that henceforth dissolves.}
    \label{fig:dolomite_3x3}
\end{sidewaysfigure}

We see the results for the miRPT model of the CDRT system in Figure \ref{fig:dolomite_3x3}.
The miRPT results are compared to an analogous FD simulation that uses operator splitting to solve the ADRE \eqref{adre} using explicit upwinding for advection, explicit centered differencing for diffusion, and PhreeqcRM for reaction calculations.
As in \cite{Leal_2016_xLMA}, the spatial distributions of pH and concentrations of calcite and dolomite are recorded and analyzed to ensure both the FD and miRPT solutions are displaying proper behavior.
We see near exact fit between the two model solutions for the times depicted in Figure \ref{fig:dolomite_3x3} ($t = 50,\ 250,\ 1000$ min), with some slight discrepancy in the position of the dissolution and precipitation fronts for calcite and dolomite at $t = 1000$.
This, however, is not particularly concerning because the difference is small, and the ``true'' solution is often unclear for systems of this type.
This is because various modeling decisions may generate solution variations on the same order of magnitude, for example, the choice of thermodynamic database, the level of discretization, or the order in which the transport and reaction operations are performed \cite{Leal_2016_xLMA,Beisman2015}.

\section{Summary and conclusions} 
\label{sec:conclusions}

In this paper, we present a Lagrangian (particle-tracking) method for modeling reactive transport with interactions between solid and aqueous chemical phases using mobile and immobile particles.
These particles transfer mass between phases (mobile to immobile, or vice versa) using an algorithm that is demonstrated to solve the diffusion equation with a controllable level of error, meaning that a high degree of accuracy may be attained for appropriate parameter choices.
This mobile-immobile reactive particle-tracking mass-transfer (miRPT MT) algorithm is physically-motivated and based on collision probabilities between particle pairs.
Mathematically, it has the form of a finite-dimensional convolution with a Green's function, which is the basis for the diffusion operator approach, derived in Section \ref{sub:derivation_of_diffusion_operator}.
However, the key difference is the double application of this convolution operator that is applied to two separate ``particle grids'' in the miRPT MT algorithm.
The regularly-spaced grid represented by the immobile particles allows miRPT MT to overcome the sometimes irregular spacing of the mobile particles (which can represent heterogeneity or reactant segregation), a problem which degraded the accuracy of the diffusion operator approach.
Additionally, we show that miRPT produces the proper concentration spatial variance increase in simulations with random mobile particle spacings, when strict convergence to an analytic solution would not be expected or an analytic solution may not be available.
This is important because, in real-world systems, global convergence is not a realistic metric, but as long as the proper dynamics are being simulated at the local (particle) level, the algorithm remains valid.
We know that the miRPT MT algorithm is simulating local diffusion dynamics because each mass transfer is a discretized convolution of the particle's spatial kernel (Dirac delta, for this work) with the diffusion Green's function, and, in the infinite particle limit, this solution is exact.

To summarize, we present the following conclusions:
\begin{itemize}
    \item The mobile-immobile reactive particle-tracking (miRPT) algorithm extends the work of \emph{Benson and Bolster} \cite{Benson_arbitrary} that enables particle-tracking methods to simulate arbitrarily complex chemical reactions by allowing particles to carry many species of reactant and transfer mass between particles.
    \item The miRPT mass-transfer (MT) algorithm allows for mass transfers between aqueous and solid (mobile and immobile) phases, enabling it to model chemical reactions such as dissolution and precipitation.
    \item The miRPT MT algorithm is demonstrated to solve the diffusion equation with a controllable level of error, depending on the level of discretization.
    \begin{itemize}
        \item For this reason, the total diffusion of a system may be partitioned between diffusive random walks and miRPT MT, allowing for the necessary mass transfers to occur without introducing spurious transport effects.
        \item Because the miRPT MT algorithm takes the form of a convolution with the diffusion Green's function, it can simulate any length of time step ($\dt$), as long as attention is given to the stability condition, discussed throughout Section \ref{sec:results}.
        \item The miRPT MT algorithm displays rapid \revtwo{(superlinear)} convergence as the number of mobile and/or immobile particles $\left(N_M, N_I\right)$ in a simulation is increased, and the solution approaches the exact solution in the limiting case.
        \item \revtwo{While this work partitioned the simulation of diffusion only between random-walking mobile particles and 2 directions of mobile-immobile mass transfers, there is no reason mobile-mobile mass transfers could not be additionally used to simulate diffusion. In fact, this may obviate the need for a large number of immobile particles, as, under the current paradigm, mobile particles that are not near any immobile particle would not simulate their full level of diffusion unless they can make diffusive mass transfers to other mobile particles.}
    \end{itemize}
    \item In the context of a full reactive transport simulation (calcite/dolomite dissolution/precipitation is considered in this work), the miRPT algorithm is demonstrated to perform effectively and generate similar solutions to an analogous Eulerian finite difference model.
\end{itemize}


\section{Acknowledgments} 
\label{sec:acknowledgments}

We thank the editor and reviewers for their helpful and incisive comments.
The first author thanks Heewon Jung and Hannah Podzorski for all of their assistance with geochemical questions.


\appendix
\section{\texorpdfstring{Analysis of the effect of choosing $\kappa^{(p)}$}{Analysis of the effect of choosing kappap}}
\label{appa}

In this section, we consider the effect of the scaling constant, $\kappa^{(p)},\ p = M, I$ in \eqref{co_loc_density}, on the behavior of the miRPT MT algorithm.
However, first we would like to show that this method of partitioning diffusion between separate mass transfers is valid.

\subsection{\texorpdfstring{Justification for partitioning diffusion via $\kappa^{(p)}$}{Justification for partitioning diffusion via kappap}} 
\label{sub:justification_for_partitioning_diffusion_via_}

\revone{
Let us first choose $\kappa^{(M)} = \alpha$ and $\kappa^{(I)} = 1 - \alpha$, $\alpha \in (0, 1)$.
We wish to show that the composed operator $\left[\vec W_I \vec W_M\right]$, as given in \eqref{wiwm}, is equivalent to an operator that simulates the total diffusion of a system, namely, $\vec D_{\dt}$, as given in \eqref{diff_op_xfer}.
Stated using the matrix form of these operators, we would like to show
\begin{equation}
    \vec D_{\dt} = \vec W_I \vec W_M,
\end{equation}
or, expressed entrywise
\begin{equation}
    \vec D_{ij} = \frac{1}{\left[4 \pi D \dt\right]^{(d/2)}}\exp\left[-\frac{\left\Vert x_i - x_j \right\Vert^2}{4 D \dt}\right] = \sum_{k = 1}^{N} \left[\vec W_I\right]_{ik} \left[\vec W_M\right]_{kj}.
\end{equation}
Next, we employ the definitions for a given entry of $\vec W_p$
\begin{equation}
\begin{aligned}
    \sum_{k = 1}^{N} \left[\vec W_I\right]_{ik} \left[\vec W_M\right]_{kj}
    &= \sum_{k = 1}^{N} \frac{1}{\left[(1- \alpha) 4 \pi D \dt\right]^{(d/2)}}
    \begin{aligned}[t]
        &\exp\left[-\frac{\left\Vert x_i - x_k \right\Vert^2}{(1- \alpha) 4 D \dt}\right] \times\\
        &\frac{1}{\left[\alpha 4 \pi D \dt\right]^{(d/2)}}\exp\left[-\frac{\left\Vert x_k - x_j \right\Vert^2}{\alpha 4 D \dt}\right]
    \end{aligned}\\
    &\defeq \sum_{k = 1}^{N} \psi_{(1 - \alpha)}(x_i - x_k) \psi_{\alpha}(x_k - x_j)
\end{aligned}
\end{equation}
Now, in order to show equality, we must consider the limiting, infinite particle, case so our sum over $k$ becomes an integral over $x_k$
\begin{equation}
    \int_{-\infty}^{\infty} \psi_{(1 - \alpha)}(x_i - x_k) \psi_{\alpha}(x_k - x_j) dx_k.
\end{equation}
If we make the substitution $x_i - x_k \to \tau$ and use the symmetry of the Gaussian, we have
\begin{equation}
    \int_{-\infty}^{\infty} \psi_{(1 - \alpha)}(\tau) \psi_{\alpha}(x_j - x_i - \tau) d\tau = \left(\psi_{(1 - \alpha)} \star \psi_{\alpha}\right)(x_j - x_i),
\end{equation}
where $\star$ denotes convolution.
Finally, we use the property that the convolution of Gaussian densities is a Gaussian density with a variance that is the sum of the individual variances and a mean that is the sum of the individual means (note that $\psi$ has mean zero).
Thus, we have
\begin{equation}
\begin{aligned}
    \left(\psi_{(1 - \alpha)} \star \psi_{\alpha}\right)(x_j - x_i) &= \psi_1(x_i - x_j)\\
    &\equiv \vec D_{ij}.
\end{aligned}
\end{equation}
}

\revone{
We note here that this justification relies on considering the infinite particle case. However, any approximation of this continuous solution will incur a controllable level of error related to the discretization but will be the best approximation in the finite-dimensional space generated by that discretization.
}


\subsection{\texorpdfstring{Relationship between error and $\kappa^{(p)}$}{Relationship between error and kappap}} 
\label{sub:relationship_between_error_and_kappap_}

Next, we look to numerically investigate the influence of the choice of $\kappa^{(p)}$ on the error of a simulation.
For these simulations, $\dt = 1.0 \times 10^{-1},\ N_M = 1000$, the total diffusion in the simulation is partitioned between $\kappa^{(M)}$ and $\kappa^{(I)}$ such that $\kappa^{(M)} = 1 - \kappa^{(I)}$, and a Gaussian IC is employed.
We consider the parameter range $\kappa^{(p)} \in \{0.1, 0.2, \dots, 0.9\}$ and test these parameters both for equally-spaced (as in Section \ref{ssub:mass_transfer_analysis_for_equally_spaced_particles}) and randomly-spaced (as in Section \ref{ssub:mass_transfer_analysis_for_randomly_spaced_particles}) mobile particles.
We conduct these tests for the range of values $N_I \in \{100, 200, 500, 1000\}$.

\begin{figure}[tp]%
    \centering
    \subfloat[RMSE vs. $\kappa^{(M)}$ (left axis) and stability condition (right axis) for $N_M = N_I = 1000$.]{\includegraphics[width=0.48\textwidth]{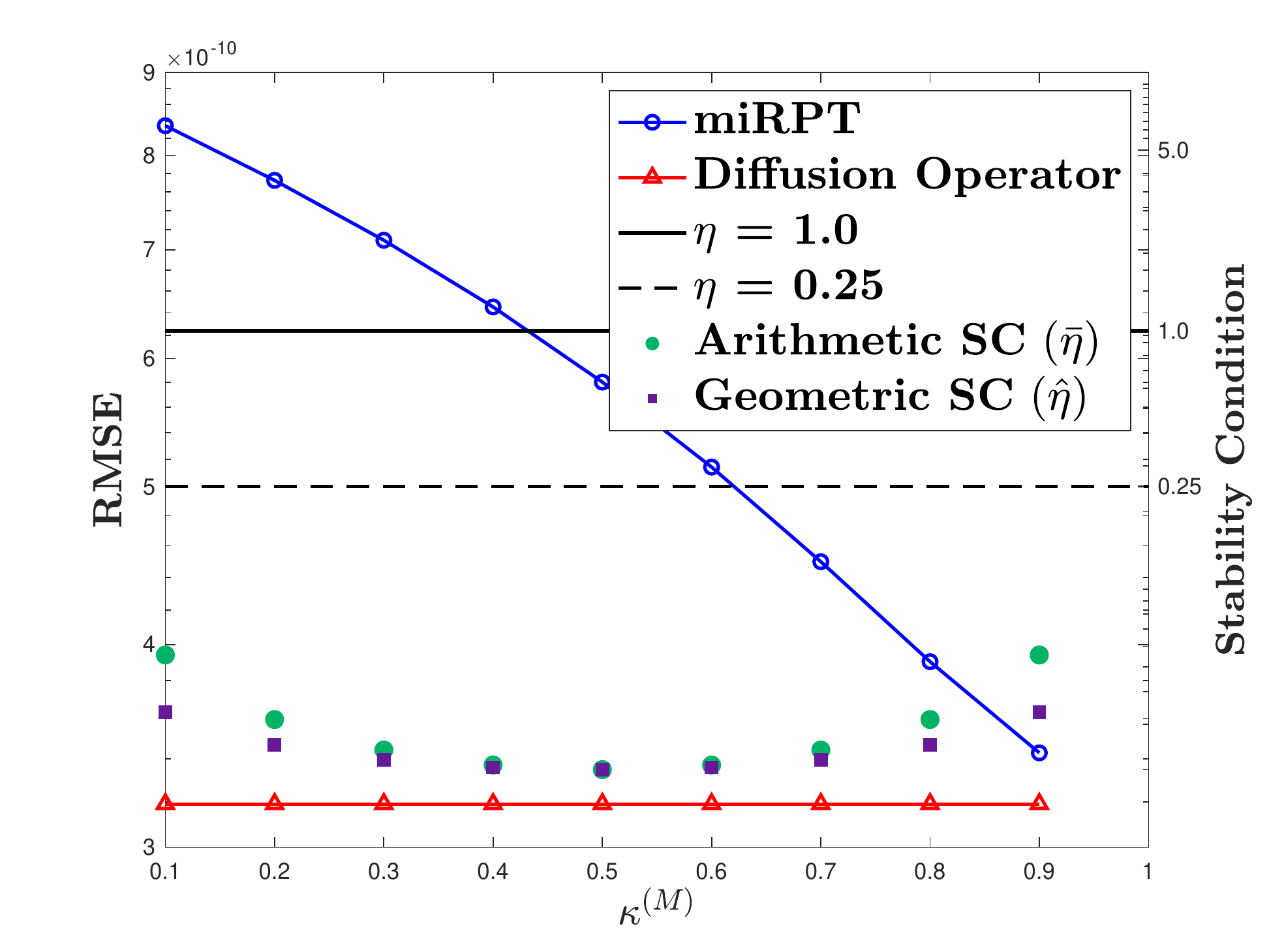}}
    \hspace{0.01\textwidth}
    \subfloat[RMSE vs. $\kappa^{(M)}$ (left axis) and stability condition (right axis) for $N_M = 1000,\ N_I = 500$.]{\includegraphics[width=0.48\textwidth]{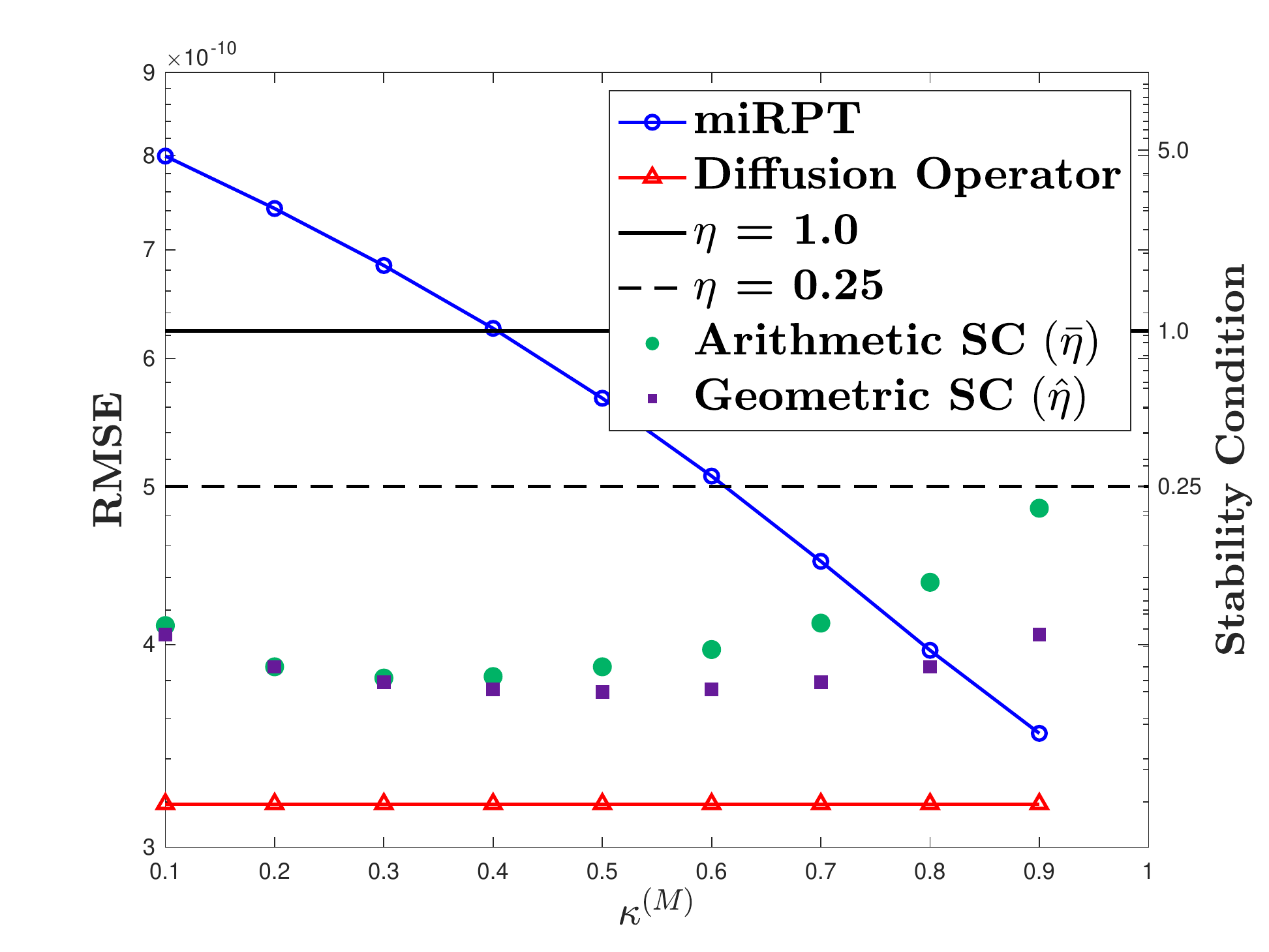} }
    \\
    \subfloat[RMSE vs. $\kappa^{(M)}$ (left axis) and stability condition (right axis) for $N_M = 1000,\ N_I = 200$.]{\includegraphics[width=0.48\textwidth]{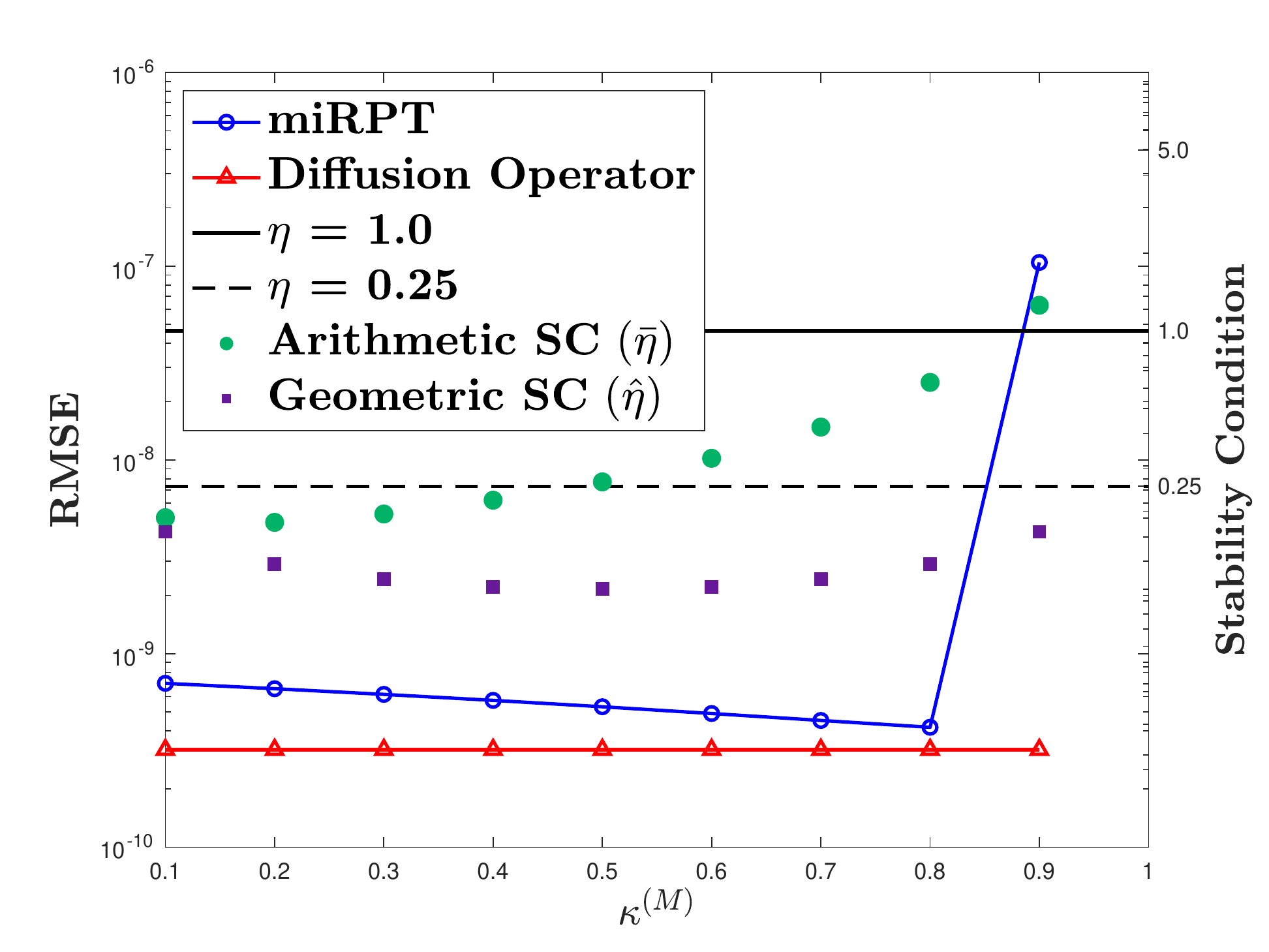}}
    \hspace{0.01\textwidth}
    \subfloat[RMSE vs. $\kappa^{(M)}$ (left axis) and stability condition (right axis) for $N_M = 1000,\ N_I = 100$.]{\includegraphics[width=0.48\textwidth]{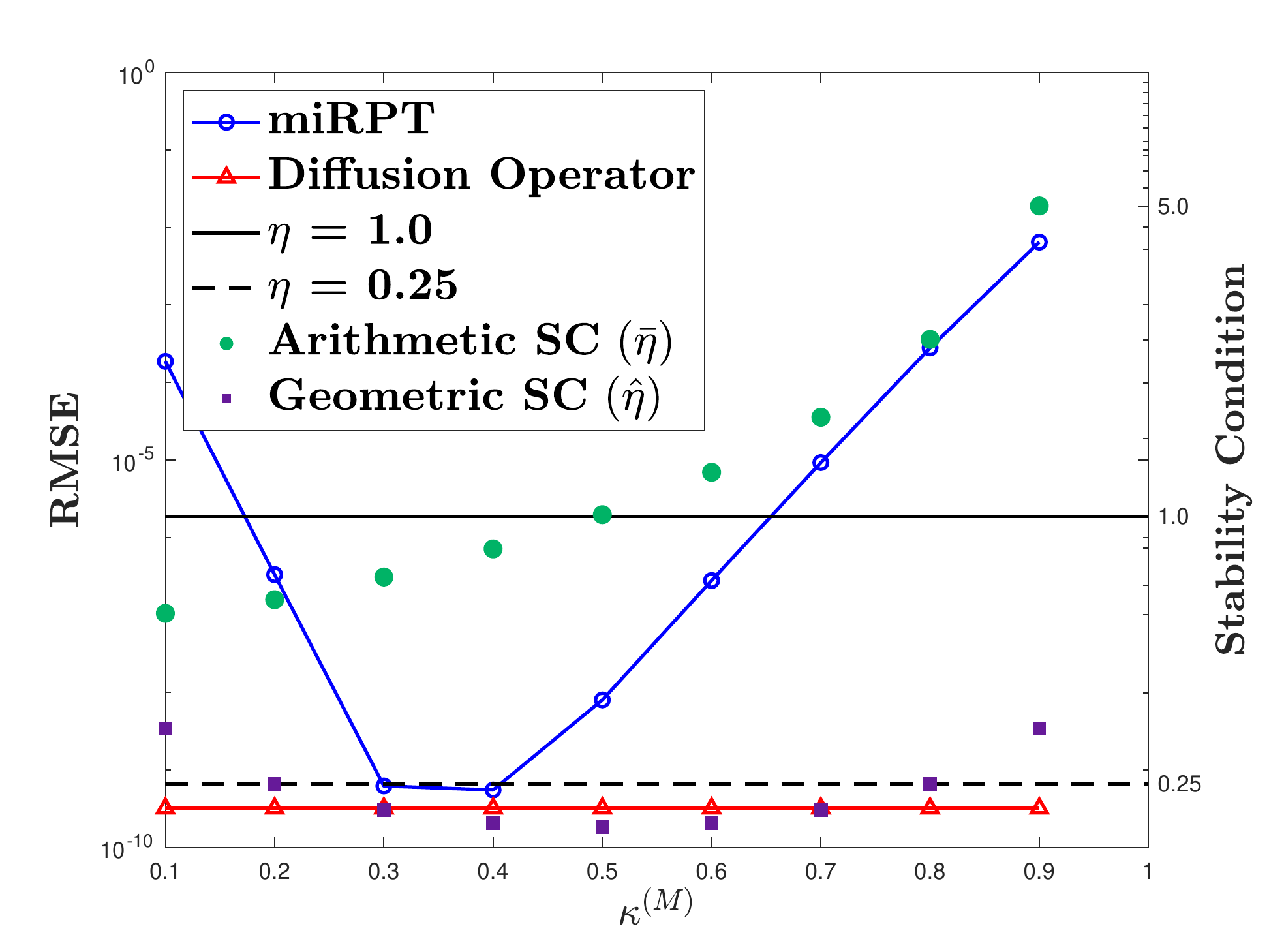} }
    \caption{Error analysis as a function of $\kappa^{(M)} \left(= 1 - \kappa^{(I)}\right)$ for Gaussian initial condition with equally-spaced mobile particles (see Figure \ref{fig:equi_dt_wGauss}(a) for reference).}
    \label{fig:equi_kappa}
\end{figure}

\begin{figure}[tp]%
    \centering
    \subfloat[RMSE vs. $\kappa^{(M)}$ (left axis) and stability condition (right axis) for $N_M = N_I = 1000$.]{\includegraphics[width=0.7\textwidth]{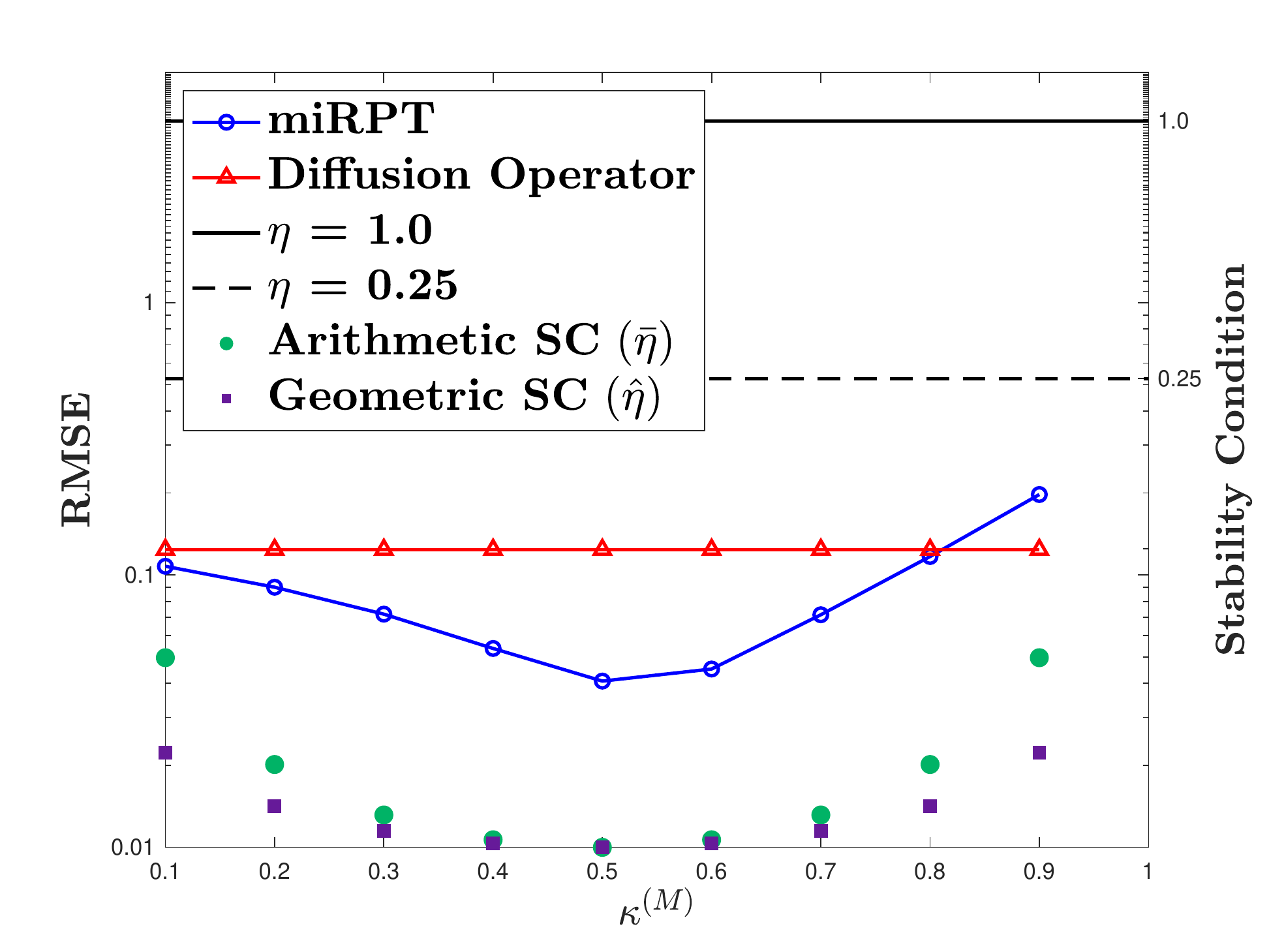}}
    \hspace{0.01\textwidth}
    \subfloat[RMSE vs. $\kappa^{(M)}$ (left axis) and stability condition (right axis) for $N_M = 1000, N_I = 100$.]{\includegraphics[width=0.7\textwidth]{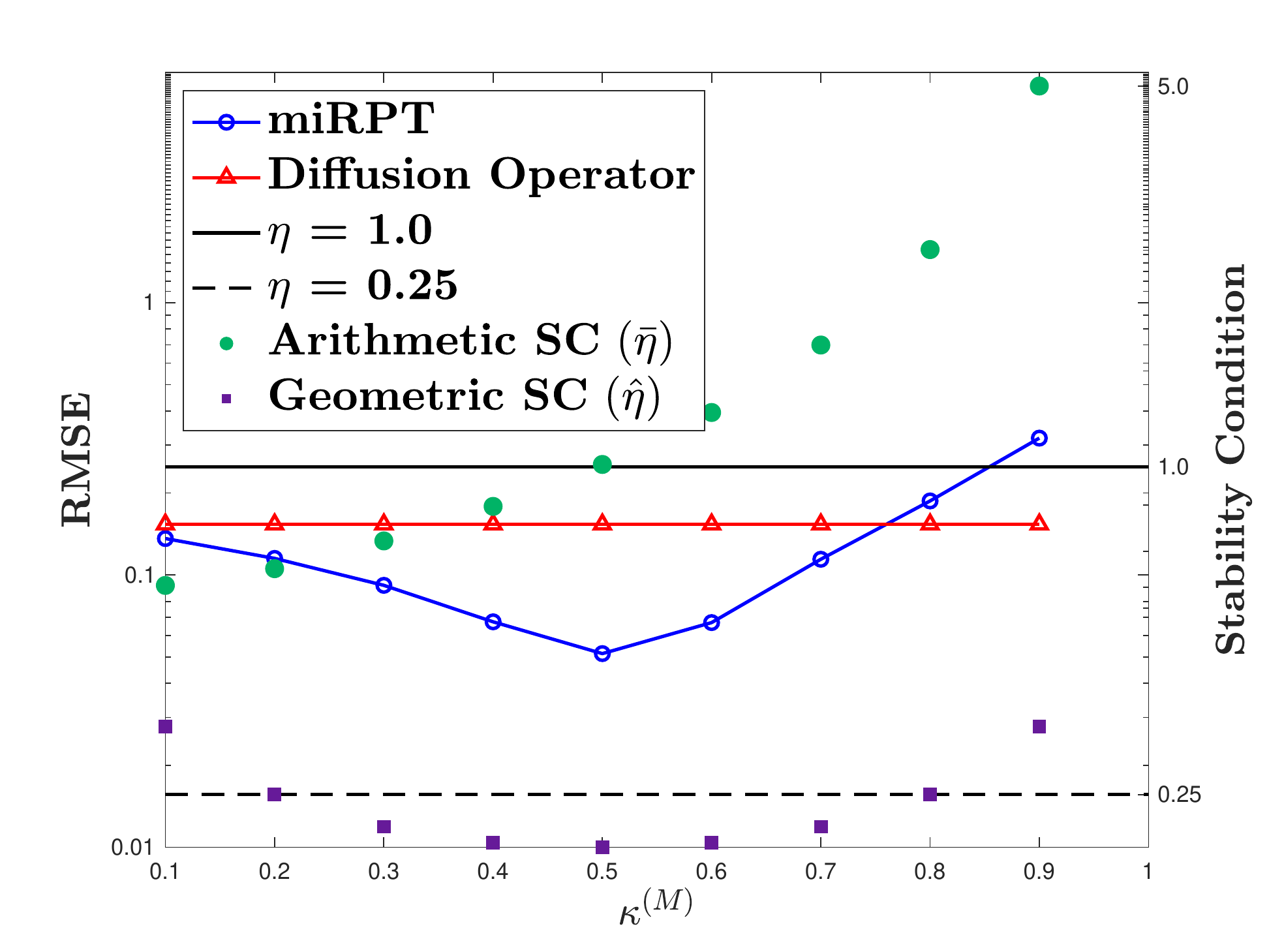} }
    \caption{Error analysis as a function of $\kappa^{(M)} \left(= 1 - \kappa^{(I)}\right)$ for Gaussian initial condition with randomly-spaced mobile particles (see Figure \ref{fig:var_gaussian}(a) for reference).}
    \label{fig:rand_kappa}
\end{figure}

For these results, we consider a generalization of the stability condition given by \eqref{eta1}
\begin{equation}
    \eta_{\kappa^{(p)}} \defeq \frac{(L/ N_p)^2}{\kappa^{(p)} D \Delta t},\quad p = M, I,
\end{equation}
leading to the condition
\begin{equation}\label{eta_avg}
    \bar \eta \defeq \frac{\eta_{\kappa^{(M)}} + \eta_{\kappa^{(I)}}}{2} \leq 1,
\end{equation}
which is the arithmetic mean of $\eta_{\kappa^{(M)}}$ and $\eta_{\kappa^{(I)}}$ (denoted \emph{arithmetic stability condition (SC)} in Figures \ref{fig:equi_kappa}-\ref{fig:rand_kappa}).
We note that, for the choice of $\kappa^{(M)} = \kappa^{(I)} = 0.5$, as specified in Section \ref{ssub:algorithm_details}, and $N_I = N_M$, we recover the form of $\eta$ given in \eqref{eta1} with the altered stability condition of $\eta \leq 1 / 2$.
However, since $\kappa^{(p)},\ p = M, I$ are treated as constants of the miPRT MT algorithmic formulation for the analysis done in Section \ref{sub:analysis_of_mobile_immobile_mass_transfer_algorithm}, we do not include it in any of those analyses.

We see in Figure \ref{fig:equi_kappa} the results for the equally-spaced mobile particle test.
The results depicted in Figure \ref{fig:equi_kappa}(a) and (b) demonstrate that, for a ratio as low as $N_I = N_M /2$, the error in the miRPT MT algorithm is relatively insensitive to the choice of $\kappa^{(p)}$, as the errors are all on the order or $1.0 \times 10^{-10}$ and comparable to the diffusion operator error.
This is not surprising, as all values of $\kappa^{(p)},\ p = M, I$ keep the simulation in the stable region of parameter space (in terms of $\bar \eta$).
However, considering Figures \ref{fig:equi_kappa}(b) and (c), we see the effects of exiting the stability region.
In Figure \ref{fig:equi_kappa}(c), where $N_M = 1000$ and $N_I = 200$, we see that the error is mostly insensitive to the values of $\kappa^{(p)}$, until the arithmetic stability condition is violated $\left(\bar \eta > 1\right)$, and we see a jump in error of approximately two orders of magnitude.
Figure \ref{fig:equi_kappa}(d) shows similar results with one key difference.
We see the characteristic large jump in error when $\bar \eta > 1$; however, we also see a sizable jump in error on the low side, before we attain the minimal level of error in the range $\kappa^{(M)} \in \{0.3, 0.4\}$.
This suggests that there is something more nuanced occurring and that perhaps the stability condition formulated above is not quite restrictive enough when the ratio $N_I / N_M$ becomes small.
We propose one possibility that involves introducing the quantity
\begin{equation}
    \hat \eta \defeq \sqrt{\eta_{\kappa^{(M)}} \eta_{\kappa^{(I)}}}\ ,
\end{equation}
which is the geometric mean of $\eta_{\kappa^{(M)}}$ and $\eta_{\kappa^{(I)}}$ (denoted \emph{geometric stability condition (SC)} in Figures \ref{fig:equi_kappa}-\ref{fig:rand_kappa}).
The geometric mean is a suitable choice for averaging these two quantities because it does not scale linearly, like the arithmetic mean, and so is able to capture the average of quantities that may have a significant difference in scale.
Using this, the stability region appears to be of the form $\left\{\bar \eta \lessapprox 1\right\} \cap \left\{\hat \eta \lessapprox 0.25\right\}$, as there does not appear to be a strict cutoff when considering $\bar \eta$ and $\hat \eta$ together.

In Figure \ref{fig:rand_kappa}, we show the results for the randomly-spaced mobile particle test.
For these simulations, mobile particle positions were assigned according to independent draws from a $\cU(0, 1)$ distribution, as in Section \ref{ssub:mass_transfer_analysis_for_randomly_spaced_particles}.
In this analysis, the magnitude of the errors is relatively large, as no ensemble-averaging (as in Section \ref{par:ensemble_run_for_gaussian_ic}) is conducted; however, the trend is clear in that there are obvious minima at $\kappa^{(M)} = \kappa^{(I)} = 0.5$ for both the $N_I = N_M = 1000$ and $N_I = N_M / 10 = 100$ cases (Figures \ref{fig:rand_kappa}(a) and (b), respectively).

Additionally, as in the ensemble-averaged results given in Section \ref{ssub:mass_transfer_analysis_for_randomly_spaced_particles} the stability condition has a different behavior for randomly-spaced particles than it does for equally-spaced.
The important result, though, is that the minimum error reliably occurs at the point where $\hat \eta$ is also minimized, and this result holds reliably for all tested ratios of $N_I / N_M$ between $0.1$ and $1.0$.
For this reason, the parameters $\kappa^{(M)} = \kappa^{(I)} = 0.5$ \revtwo{are held constant} for all tests conducted in Section \ref{sec:results}.

\section{Supporting information for reactive transport system} 
\label{sec:supporting_information_for_reactive_transport_system}

\subsection{Selection of model parameters} 
\label{sub:selection_of_model_parameters}

As a starting point, we first ensure that the FD model replicates the results of the CDRT system presented in \cite{Leal_2016_xLMA}.
This led to choosing a time step of $\dt = 1.0$ seconds and a spatial discretization of $\dx = 1.0 \times 10^{-3}$ m, which results in a Courant number of $C_{FD} \defeq (v \dt) / \dx = 0.024$ and a grid P\'eclet number of $Pe_{FD} \defeq (v \dx) / D = 0.2$.
We note here that the choice of $\dt$ requires consideration also be given to the timescale of the chemical reactions being modeled, as the equilibrium reactions are considered to be instantaneous.
However, that is not the primary focus of this text, so we merely verify that the FD model reproduces the original results of \emph{Leal et al.} and proceeded to use that value of $\dt$ in our miRPT model as well.

The next step is to choose the proper level of spatial discretization (i.e., the number of mobile and immobile particles, $N_M$ and $N_I$), given the the stability constraint provided by the miRPT MT algorithm and the defined parameters of the CDRT system, $D = 1.2 \times 10^{-7}$, $\Omega = 0.5$, along with the choice of $\dt = 1.0$.
For our miRPT model of the CDRT system, we choose to simulate half of the total diffusion by random walks and half using the miRPT MT algorithm (for this reason we impose $D_{MT} = 0.6 \times 10^{-7}$ for the following miRPT MT analysis).
This 50/50 partitioning of diffusion is somewhat arbitrary in this case, though the analysis would be similar regardless of this modeling choice because the value of the stability parameter, $\eta$, stays on the same order of magnitude for any partition that sums to unity.
A discussion of employing this partition between random walk and MT diffusion in order to differentiate between macro-scale dispersion and micro-scale mixing may be found in \cite{mass_trans_acc}.

\begin{figure}[tp]%
    \centering
    \subfloat[Spatial variance increase vs. time for $N_I = 4000,\ N_M = 5000$.]{\includegraphics[width=0.7\textwidth]{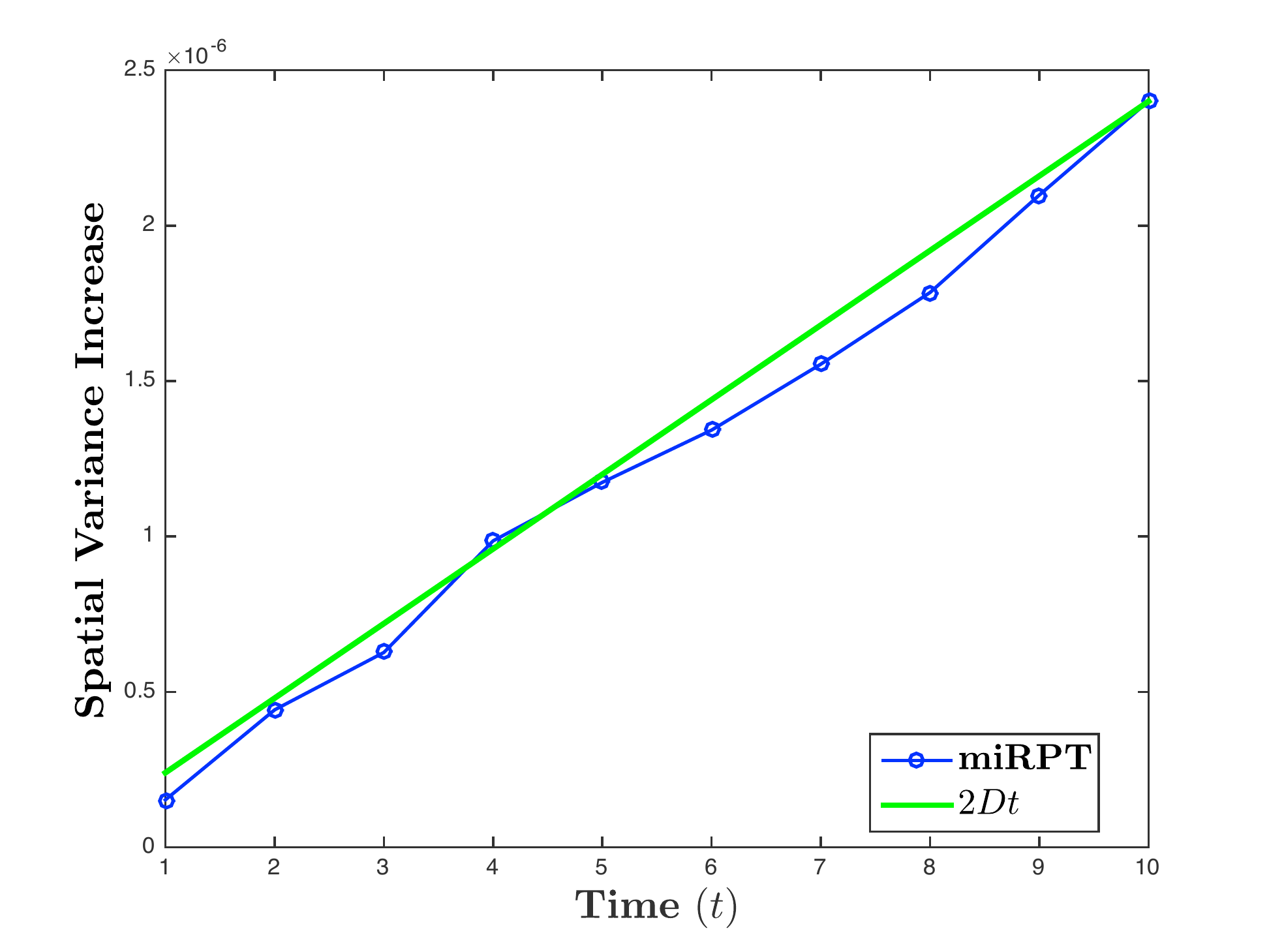}}
    \hspace{0.01\textwidth}
    \subfloat[Spatial variance increase vs. time for $N_I = 3000,\ N_M = 6000$.]{\includegraphics[width=0.7\textwidth]{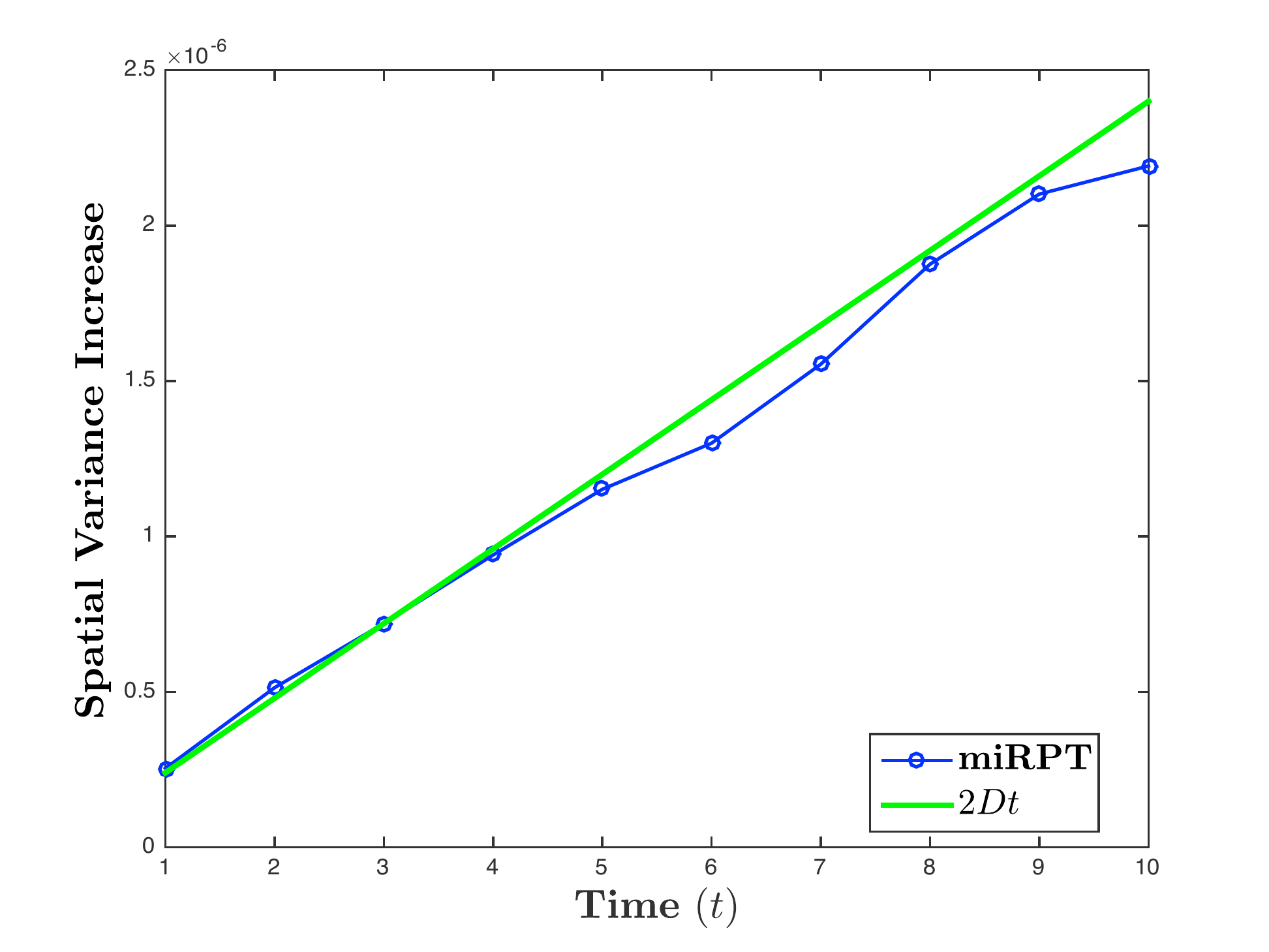}}
    \caption{Error analysis for calcite-dolomite system, in terms of variance increase with Dirac delta initial condition and random-walking mobile particles.}
    \label{fig:dolomite_RW_var1}
\end{figure}

The first step we take is to verify that the miRPT MT algorithm is capable of simulating the low level of diffusion in the CDRT for stationary mobile particles in both the equally- and randomly-spaced cases.
The results of this analysis are not shown here, as they are positive and mirror those seen in Section \ref{sub:analysis_of_mobile_immobile_mass_transfer_algorithm}.
Next, we consider spatial variance increase in the case where the mobile particles are simulating diffusion via random walks.
These conditions capture the total action (in terms of variance increase) of the conservative transport portion of the CDRT system, \revtwo{as homogeneous advection simulation should not induce variance increase}.
This is, of course, ideally speaking, as FD schemes for advection simulation are known to induce numerical diffusion, a feature which is not present in particle-tracking methods.
The results shown in Figure \ref{fig:dolomite_RW_var1} for a Dirac delta IC (10 second simulations are depicted, though results are nearly identical for longer periods of time) demonstrate very close agreement between expected and simulated variance increase for $N_I = 4000,\ N_M = 5000$ ($\eta \approx 0.26$) in Figure \ref{fig:dolomite_RW_var1}(a) and $N_I = 3000,\ N_M = 6000$ ($\eta \approx 0.46$) in Figure \ref{fig:dolomite_RW_var1}(b).
The maximum and final error in both cases is nearly two orders of magnitude smaller than the value of the variance, itself.
This reinforces the previously mentioned assertion that the random walk diffusion for mobile particles works to smooth the effects of any irregular spatial position distributions.
An important result here is that a larger number of particles (both mobile and immobile) is required to adequately model the desired level of diffusion than in the cases without random walks (satisfactory results were achieved using half as many particles for some of the stationary particle cases), indicating that a good rule of thumb may be to keep $\eta < 0.5$.
Additionally, for both $N_I = 4000,\ N_M = 5000$ and $N_I = 3000,\ N_M = 6000$ there is a significant increase in error if $N_I$ is reduced much below these values or if $N_M$ is reduced much below 5000.

From an accuracy perspective, either of the parameter pair choices from the random walk/miRPT MT simulations, shown in Figure \ref{fig:dolomite_RW_var1}, would be appropriate.
As well, from a computational perspective, if we are considering a conservative system (i.e., without chemical reactions), the choice of $N_I = 4000,\ N_M = 5000$ may be superior.
This is because, while both choices result in 9000 total particles (leading to approximately the same number of total mass-transfer calculations), choosing to employ fewer mobile particles will lead to fewer transport calculations and a faster simulation.
However, the primary computational cost, by far, in the CDRT system is due to the chemical reaction calculations that are performed by PhreeqcRM.
So, we would like to minimize the number of these calculations that are performed by choosing the parameter pair with the minimum appropriate number of immobile particles.
For that reason, the miRPT CDRT model will use $N_I = 3000,\ N_M = 6000$.

\subsection{Computational details for the reactive transport system} 
\label{sub:computational_details_dolo}

In the preceding sections, we have laid the theoretical groundwork necessary for developing a reactive transport model using the miRPT algorithm.
However, from an application standpoint, actually writing the code for such a model involves some computational choices that may not be readily apparent.
For this reason, we intend this section to act as something of a high-level ``user's guide'' for the miRPT model.
Additionally, we make our CDRT FD and miRPT code available at \texttt{doi.org/10.5281/zenodo.2558584} \cite{miRPT_code}.

We begin the CDRT simulation by initializing particle positions.
Immobile particles are equally-spaced on the interval $\Omega = [0.0, 0.5]$ (this may be thought of as creating an Eulerian grid with $\dx = 1 / N_I$), and mobile particles are assigned positions according to draws from a $\cU(0.0, 0.5)$ distribution.
Once the geochemical solver, PhreeqcRM, is initialized, we pass our IC (calcite and quartz within the domain) and BC concentrations (CO$_2$ and salts to be injected) to it, so that we may assign initial concentrations of our solid phases and aqueous ions to the appropriate particles.
All of the immobile particles are assigned the concentrations generated from the IC.
The BC is slightly more complicated in the particle context.
In the FD context, the ``constant concentration injection'' BC may be interpreted as a constant-flux, or Neumann BC.
To match this type of BC, concentrations are assigned to the lower-boundary immobile particle ($x = 0.0$).
Once time-stepping and transport begin, negative fluxes at the lower boundary are prevented by imposing a ``reflecting'' BC on the diffusive random walks of the mobile particles (advection only moves in the positive direction for this problem).
So, a particle that randomly walks across the lower boundary and attains a negative $x$-position is re-assigned the absolute value of that $x$-position.

With the simulation initialized, we convert the concentrations (mol/L) of the aqueous species on the immobile particles (the species undergoing transport) to masses (mol/particle) by multiplying by the representative volume of an immobile particle $\left(V_0 = \left(0.5 \text{\ m} \times 1000 \text{\ L/m}\right) / N_I\right)$ (step (f) in Figure \ref{fig:imRPT_chart}).
These masses are then transferred to the mobile particles according to the miRPT MT algorithm (step (g) in Figure \ref{fig:imRPT_chart}, details described in the following section).
Here, time-stepping begins, and random-walk diffusion and advection are simulated sequentially by altering mobile particle positions according to \eqref{transport_langevin} (step (b) in Figure \ref{fig:imRPT_chart}).
A minor computational note regarding the random-walk algorithm is that, as this code was written in Fortran, there is not a built-in normally-distributed random number generator, so we use the Box-Muller transformation \cite{box_muller_og} to generate $\cN(0, 1)$ random numbers using $\cU(0, 1)$ random numbers (for example code, see \cite{miRPT_code,box_muller_code_site}).
Next, BCs are enforced according to the previously described reflecting lower boundary, as well as an ``absorbing'' upper boundary ($x = 0.5$) that can be interpreted as a homogeneous Dirichlet BC in the FD context.
Computationally, this absorbing boundary is enforced by reassigning any mobile particle that crosses the upper boundary to have position $x = 0.0$ and carry no mass.
This ``wraparound'' reassignment to the lower boundary allows the simulation to maintain a constant particle number.
Next, the masses of the aqueous species on the mobile particles are transferred to the immobile particles via the miRPT MT algorithm and converted to concentrations (steps (c) and (d) in Figure \ref{fig:imRPT_chart}).
These concentrations are passed to PhreeqcRM for the chemical reaction calculations (step (e) in Figure \ref{fig:imRPT_chart}).
After reaction, we apply the injection BC again, and then proceed to steps (f) and (g) in Figure \ref{fig:imRPT_chart} and repeat the steps in Figure \ref{fig:imRPT_chart} until we reach the final simulation time.
We note here that the BC injection scheme we employ allows for a constant flux of aqueous species mass through the boundary even if no mobile particles leave the upper boundary (and then enter the lower boundary) because the BC is injected into the lower-boundary \emph{immobile} particle in every time step.
That is to say, assigning the BC mass to an immobile particle ensures the proper mass flux in each time step, as injecting the mass on mobile particles (that may or may not enter the lower boundary in a given time step) would introduce an undesirable random variability to the BC.

\subsubsection{Computational details for the mass-transfer process} 
\label{ssub:computational_details_for_the_mass_transfer_process}

In order to conduct the miRPT MT calculation, two pieces of information are required: the positions and masses of aqueous species on all mobile and immobile particles ($\vec x^{(p)}$ and $\vec m^{(p)}$, respectively, for $p = M, I$).
Using the position vectors, $\vec x^{(M)}$ and $\vec x^{(I)}$, we construct a pairwise distance matrix, $\vec \Delta$, where $\vec \Delta_{i, j} = \left\vert x_i^{(I)} - x_j^{(M)} \right\vert$.
These entries of $\vec \Delta$ are calculated using a kD tree algorithm \cite{og_kd,matlab_kd,kennel} to perform a fixed-radius search and find the nearest neighbors within a specified cutoff distance (this is discussed in further detail in Section \ref{ssub:computational_details}).
It is worth noting that $\vec \Delta$ may be constructed in sparse fashion, which can often lead to significant computational speedup and memory savings if sparse linear algebra methods are employed.
We then use $\vec \Delta$ to construct the relevant MT matrix ($\vec{W_p},\ p = M, I$, depending on the direction of transfer) in the following way, using the weighting function given in \eqref{co_loc_density} for $\kappa^{(M)} = \kappa^{(I)} = 0.5$
\begin{align}
    \vec {\tilde W_M} &= \exp\left(\frac{\vec \Delta}{- 2 D_{MT} \dt}\right),\label{comp_wt_mat1}\\
    \vec{W_M} & = \vec{\tilde W_M} \diag^{-1}\left(\vec 1 \vec{\tilde W_M}\right),\label{comp_wt_mat2}\\
    \vec{\tilde W_I} &= \exp\left(\frac{\vec \Delta^T}{- 2 D_{MT} \dt}\right),\label{comp_wt_mat3}\\
    \vec{W_I} & = \vec{\tilde W_I} \diag^{-1}\left(\vec 1 \vec{\tilde W_I}\right),\label{comp_wt_mat4}
\end{align}
where exponentiation is considered to be an element-wise operation, and $D_{MT}$ is the diffusion constant corresponding to the portion of diffusion to be simulated by miRPT MT.
Also, $\vec 1$ is a $1 \times N$ vector of ones, and $\diag^{-1}\left(\vec x\right)$ is a square matrix with the element-wise reciprocal of the vector $\vec x$ on its main diagonal, as employed previously.
We also note that \eqref{comp_wt_mat1} and \eqref{comp_wt_mat3} do not include multiplication by the constant term in \eqref{co_loc_density} because it would divide out due to the column-normalization step in lines \eqref{comp_wt_mat2} and \eqref{comp_wt_mat4}.
Finally, the mass-transfers are computed according to \eqref{matmul}, where for a immobile to mobile transfer, the formula would be $\vec W_I \vec m^{(I)} = \vec m^{(M)}$.





\bibliography{mobile_immobile_JCP}

\end{document}